\documentclass[lettersize,journal]{IEEEtran}
\usepackage{amsmath,amsfonts}
\usepackage{algorithm}
\usepackage[noend]{algpseudocode}
\usepackage{eucal}
\usepackage{array}
\usepackage[caption=false,font=normalsize,labelfont=sf,textfont=sf]{subfig}
\usepackage{textcomp}
\usepackage{stfloats}
\usepackage{url} 
\usepackage{verbatim}
\usepackage[hidelinks]{hyperref} 
\usepackage{graphicx}
\usepackage{cite}
\usepackage{pifont}
\usepackage{makecell}
\usepackage{adjustbox} 
\usepackage{comment}
\usepackage{tikz}
\usepackage{mathabx}

\DeclareMathOperator*{\argmin}{arg\,min}
\usepackage{multirow}
\usepackage{mathrsfs}

\usepackage{algorithmicx,algpseudocode,algorithm,algpseudocode,amsmath,amssymb}

\algdef{SE}[DOWHILE]{Do}{doWhile}{\algorithmicdo}[1]{\algorithmicwhile\ #1}%

\newcommand\checkmarkn[1][]{%
  \tikz[scale=0.4,#1]{\fill(0,.35) -- (.25,0) -- (1,.7) -- (.25,.15) -- cycle;}%
}

\def\eg{\emph{e.g.}}

\def\ie{\emph{i.e.}}

\begin{document}


\title{IsoSched: Preemptive Tile Cascaded Scheduling of Multi-DNN via Subgraph Isomorphism}

\author{
Boran Zhao\IEEEauthorrefmark{1},
Zihang Yuan\IEEEauthorrefmark{1},
Yanbin Hu\IEEEauthorrefmark{1},
Haiming Zhai\IEEEauthorrefmark{1},
Haoruo Zhang\IEEEauthorrefmark{2},
Wenzhe Zhao\IEEEauthorrefmark{3},~\IEEEmembership{Member,~IEEE,}\\
Tian Xia\IEEEauthorrefmark{3},~\IEEEmembership{Member,~IEEE,}
and Pengju Ren\IEEEauthorrefmark{3},~\IEEEmembership{Member,~IEEE} \vspace{-9mm}
\thanks{\IEEEauthorrefmark{1}Boran Zhao, Zihang Yuan, Yanbin Hu and Haiming Zhai are with the School of Software Engineering.}
\thanks{\IEEEauthorrefmark{2}Haoruo Zhang is with Nanyang Technological University, Singapore.}
\thanks{\IEEEauthorrefmark{3}All other authors are with the National Key Laboratory of Human-Machine Hybrid Augmented Intelligence, National Engineering Research Center for Visual Information and Applications, and Institute of Artificial Intelligence and Robotics, Xi'an Jiaotong University, Xi’an, Shaanxi, China.}
\thanks{E-mail: pengjuren@xjtu.edu.cn (Corresponding Author).}
}

\markboth{Journal of \LaTeX\ Class Files,~Vol.~14, No.~8, April ~2023}%
{Shell \MakeLowercase{\textit{et al.}}: A Sample Article Using IEEEtran.cls for IEEE Journals}

\maketitle


\begin{abstract}

Deploying deep neural network (DNN) accelerators with Layer Temporal Scheduling (LTS) often incurs significant overheads (\eg, energy and latency), as intermediate activations must be cached in DRAM. To alleviate this issue, Tile Spatial Scheduling (TSS) has been proposed, which reduces such costs by fragmenting inter-layer data into smaller \textit{tiles} communicated via on-chip link.
However, many emerging applications require concurrently executing multiple DNNs with \textit{complex topologies}, where certain critical tasks must preempt other tasks to meet stringent latency requirements (\eg, in autonomous driving, obstacle detection must be completed on the order of tens of milliseconds). Unfortunately, existing TSS works lack support for preemptive execution, while prior preemption schemes are based on LTS and therefore inherit its costly overheads.
This observation highlights the necessity of preemptive and efficient TSS-based frameworks. Nevertheless, designing such a system presents significant challenges, particularly due to the requirement of preemption in graphs with \textit{complex topologies} (\eg, in modern large language models, the number of graph edges can scale to tens of thousands).

To tackle this challenge, we introduce IsoSched, the first framework that enables preemptive multi-DNN scheduling on TSS architecture. IsoSched  firstly formulates preemptive scheduling of graphs with \textit{complex topologies} as an integer-linear program (ILP) and a subgraph isomorphism problem; secondly, we apply Layer Concatenate and Split (LCS) for load balancing of tile pipeline; thirdly, to enable efficient subgraph matching, we integrate an Ullmann-based algorithm enhanced with Monte Carlo Tree Search (MCTS) to accelerate the search process, while leveraging compact matrix encoding (\ie, Compressed Sparse Row, CSR) to reduce memory overhead. 
IsoSched outperforms existing LTS-PRM approaches (\ie, PREMA, Planaria, CD-MSA, MoCA) in Latency-Bound Throughput (LBT), speedup, and energy efficiency, and achieves higher critical task satisfaction rates than TSS-NPRM (\ie, HASP) across varying task complexities.
\end{abstract}

\begin{IEEEkeywords}
Multi-CNN, scheduling, subgraph isomorphism.
\end{IEEEkeywords}

\section{Introduction} 



\begin{figure}[!t] 
    \centering
    \includegraphics[width=1.0\columnwidth]{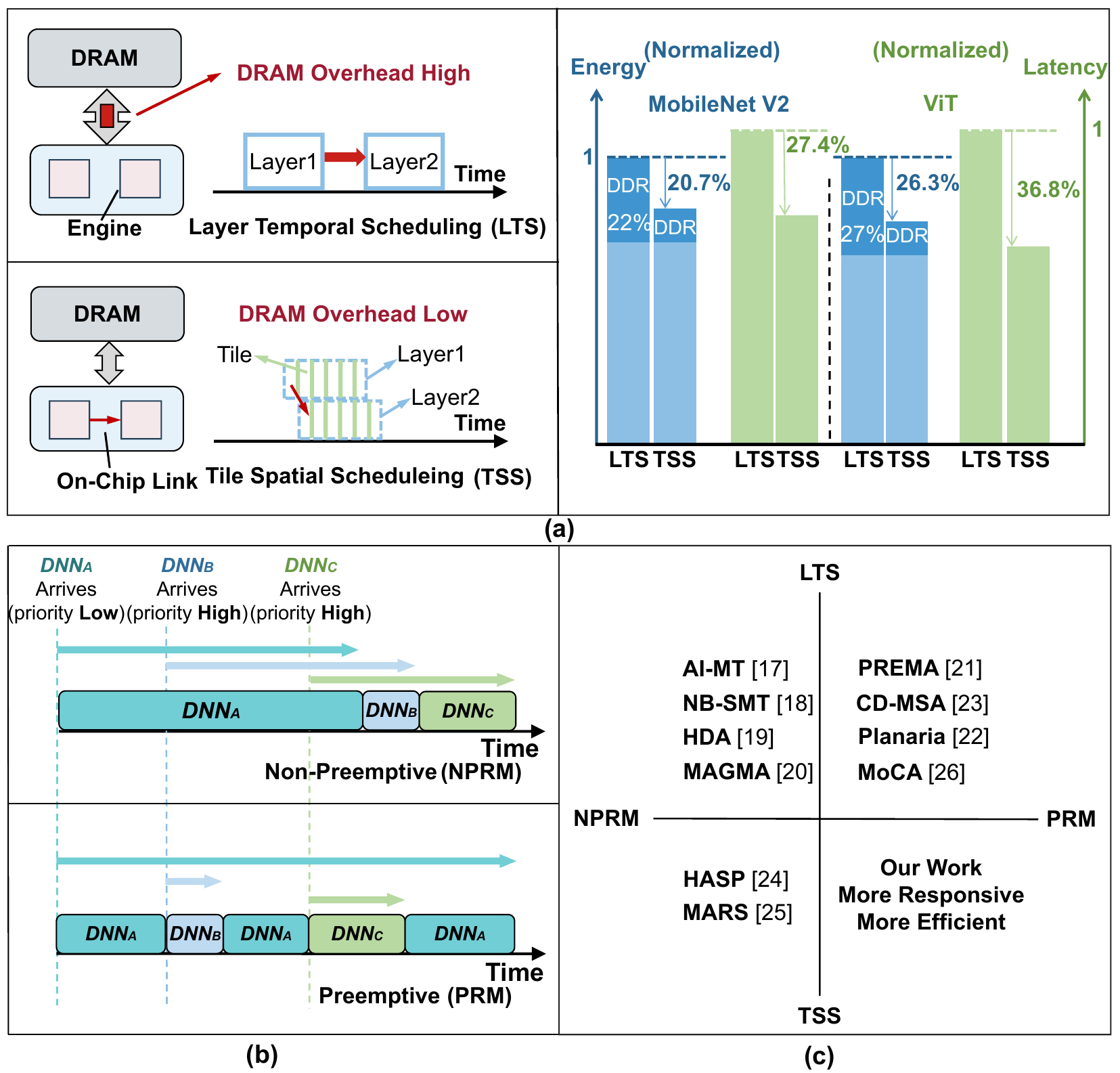}
    \caption {
    (a) Comparison of single-DNN accelerators: Tile Spatial Scheduling (TSS) outperforms Layer Temporal Scheduling (LTS) in energy efficiency and latency \cite{maestro}. (b) Comparison of multi-DNN accelerators: Preemptive (PRM) designs better accommodate high-priority tasks than Non-Preemptive (NPRM) designs. (c) Classification of prior studies and our work.
    }
    \label{fig:fig1}
\end{figure}

\IEEEPARstart{D}{eep} neural network (DNN) algorithms are typically represented as computational graphs consisting of cascaded layers~\cite{GraphRef1,GraphRef2,GraphRef3,GraphRef4,GraphRef5,zhao2025nms}. In most single-DNN accelerators, inter-layer activations are temporarily stored in off-chip DRAM, a strategy commonly referred to as Layer Temporal Scheduling (LTS)\cite{Eyeriss,NVDLA,ShiDianNao}. However, this design choice incurs substantial energy overhead due to frequent DRAM accesses—accounting for up to 27\% of the total energy consumption as shown in Fig.~\ref{fig:fig1}(a)—and results in considerable latency.

To mitigate these inefficiencies, Tile Spatial Scheduling (TSS) has been proposed~\cite{fusedlayer, remap, isos}. TSS partitions intermediate inter-layer results into smaller \textit{tiles}, which are then communicated through low-latency on-chip links. This tiling-based strategy not only substantially reduces DRAM access energy but also enables downstream layers to start computation as soon as partial outputs become available, thereby significantly reducing end-to-end inference latency.

\begin{figure}[!t] 
    \centering
    \includegraphics[width=1.0\columnwidth]{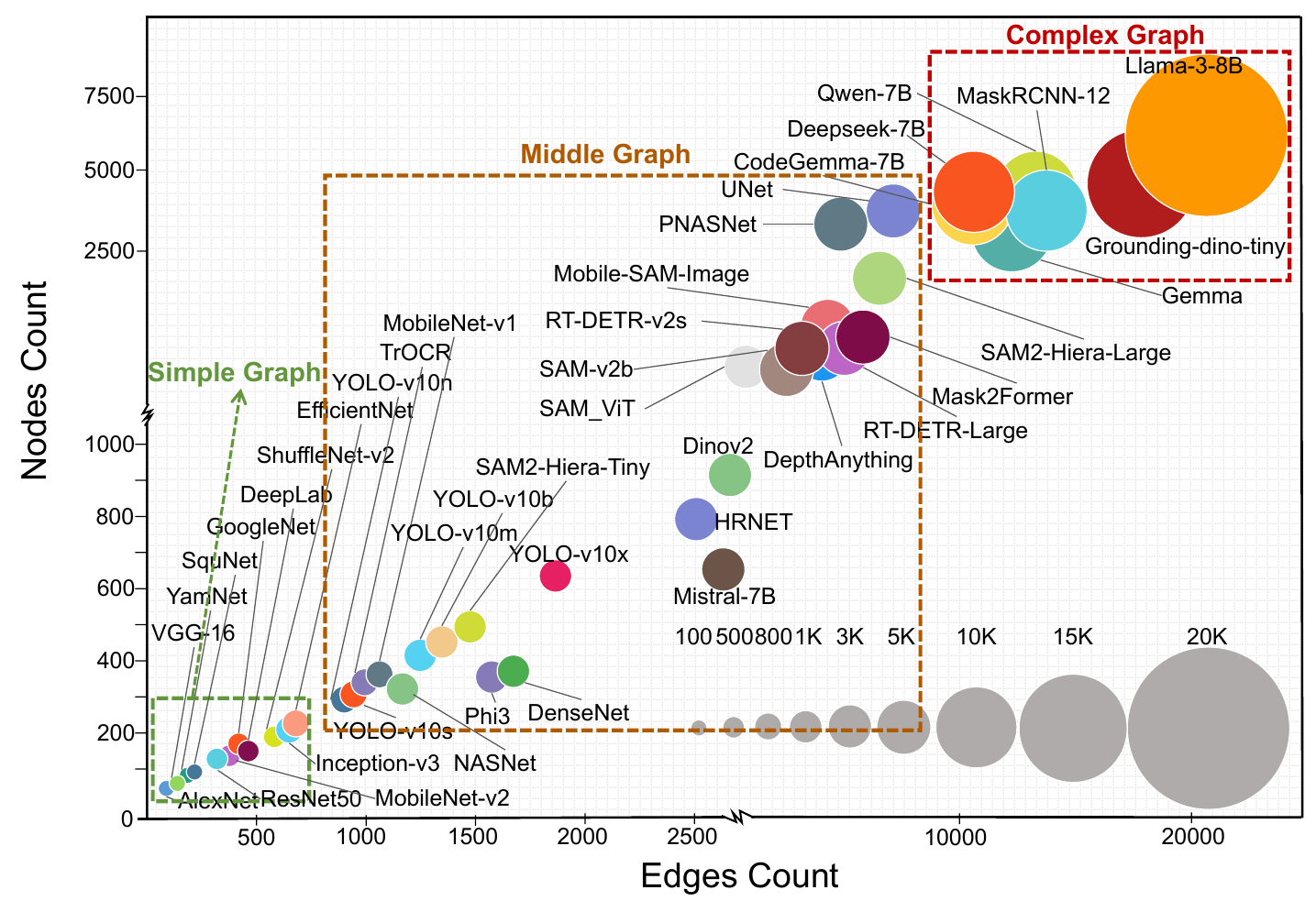}
    \caption {DAG topology complexity of different DNN tasks. Modern large language models exhibit significantly higher topological complexity.}
    \label{fig:Task_scale}
\end{figure}

\label{sec:motivation}

Beyond single-DNN execution, many real-world scenarios require concurrently running multiple DNN tasks (\ie, multi-DNN) with varying priorities on a shared accelerator. For example, autonomous driving systems must perform obstacle detection, lane segmentation, and traffic light/sign recognition simultaneously, where obstacle detection is typically assigned the highest priority~\cite{autocar_re0,autocar_re1}. In LLM-driven robotic systems, more complex task topologies often require distinct LLMs to be concurrently assigned to task decomposition, assignment, problem-solving, and result synthesis~\cite{AutoMisty}, with problem-solving typically assigned the highest priority.

To improve system throughput, several non-preemptive scheduling schemes based on LTS have been proposed for multi-DNN execution~\cite{AI-MT,NB-SMT,HDA,magma}, as illustrated in the upper-left of Fig.~\ref{fig:fig1}(c). However, these approaches are unable to ensure the timely execution of high-priority tasks.

To address this limitation, preemptive schemes under LTS have been developed. For instance, PREMA~\cite{PREMA} and Planaria~\cite{planaria} enforce preemption to prioritize critical tasks on monolithic and multi-subarray architectures, respectively. Likewise, CD-MSA~\cite{CD-MSA} guarantees deadline satisfaction for critical tasks through preemptive execution. However, these methods inevitably inherit the fundamental inefficiencies of LTS, particularly the high latency and energy overhead associated with frequent DRAM accesses.

To overcome this problem, HASP~\cite{HASP} and MARS~\cite{MARS} adopt TSS to improve overall system efficiency by utilizing on-chip cross-engine links. Despite these advantages, such methods lack preemptive support, leading to scenarios where non-critical tasks monopolize hardware resources and critical tasks fail to meet their deadlines. These insights reveal a pressing demand for preemptive support within TSS architectures.
However, to the best of our knowledge, no existing work supports preemptive multi-DNN scheduling under the TSS paradigm. This is primarily because preemptive scheduling under TSS must account not only for the allocation of computational and memory resources, but also for the reconfiguration of complex on-chip communication topologies—namely, the data links connecting processing engines. Such reconfiguration requires performing task graph matching and substitution. Moreover, the scheduling challenge is significantly amplified by the drastic increase in the complexity of modern DNN computation graphs. Compared to earlier DNNs—such as the example in Fig.~\ref{fig:Task_scale} with only hundreds of nodes and a few hundred edges—SOTA DNNs exhibit topologies with several orders of magnitude more complexity, reaching up to thousands of nodes and tens of thousands of edges.

To address these challenges, we propose IsoSched, a preemptive scheduling framework based on subgraph isomorphism. IsoSched formulates the mapping and scheduling task as an integer linear programming (ILP) problem, employs DAG-to-Pipeline (D2P) to convert the DNN DAG into a tile pipeline, and applies Layer Concatenate and Split (LCS) to balance the pipeline workload. To enable efficient subgraph matching for preemptive scheduling, the framework integrates \underline{M}onte Carlo Tree Search (MCTS) to accelerate the search process, \underline{C}ompressed Sparse Row (CSR) encoding to reduce memory overhead, and \underline{U}llmann’s algorithm as the underlying matching foundation, collectively forming the \textit{MCU} subgraph isomorphism algorithm.


The major contributions of this work are as follows:

\begin{itemize}
\item \textbf{First Preemptive Scheduler for TSS Architectures.} To the best of our knowledge, this is the first work to support preemptive multi-DNN scheduling under TSS architectures. It ensures the timely execution of critical DNN tasks while maximizing overall system throughput.
\item \textbf{Unified Scheduling Formulation for Preemptive TSS.} We propose a unified formal formulation for the preemptive scheduling of multi-DNN workloads targeting tile pipeline dataflow accelerators.
\item \textbf{Ullmann-Based Scheduling Enhanced with MCTS and Compact Matrix Encodings.} To support preemptive scheduling across multiple task graphs with complex topologies, we develop a subgraph isomorphism-based scheduling algorithm leveraging the Ullmann algorithm. To reduce the computational overhead of subgraph matching, we further incorporate MCTS and compact matrix encoding schemes, significantly reducing scheduling overhead.
\item \textbf{Layer Concatenate and Split for Pipeline Balancing.} We introduce LCS to balance tile pipelines by merging fine-grained operators and partitioning oversized ones, thereby reducing pipeline bubbles and ultimately enhancing latency-bound throughput and energy efficiency.

\end{itemize}

\section{Background and Motivation}

\label{sec:background_arch}

\subsection{Tile definition}
\label{Tile def}
\begin{equation}
\label{TileCombined}
T =
\begin{cases}
\adjustbox{max width=.85\linewidth}{$
\displaystyle
\left\lceil \frac{W_o \times C_o \times K_h \times K_w \times C_{in}}{\#\mathrm{PE_{engine}}} \right\rceil
+ \mathrm{filling\_time}
$} \\[15pt]
\adjustbox{max width=.85\linewidth}{$
\displaystyle
\left\lceil 
\frac{N_k \times H \times d_k}{\#\mathrm{PE_{engine}}}
\right\rceil
+ \mathrm{filling\_time}
$}
\end{cases}
\end{equation}

In Eq.~(\ref{TileCombined}), for convolution, $W_o$ denotes the output featuremap width, $C_o$ the number of output channels, $K_h\times K_w$ the kernel size (height and width), $C_{in}$ the number of input channels, and $\#PE_{\text{engine}}$ the number of processing elements (PE) per engine; filling\_time is the pipeline latency from the first input fed to the first output produced. For matrix multiplication in multi-head attention, $N_k$ is the number of keys (the sequence length of $K$, i.e., the width of $QK^{\top}$), $h$ the number of heads, and $d_k$ the per-head query/key dimension (the reduction size); under the \textit{tile} definition of “one output row across all heads”, the MACs per \textit{tile} equal $N_k h d_k$. For all compute-bearing DNN layers and matrix multiplication layers, we evaluate their corresponding $T$ values using Eq.~\eqref{TileCombined}, and select the minimum as the base tile time unit. This minimum $T$ defines the fundamental scheduling granularity—referred to as the engine \textit{timeslot}—used in engine-level scheduling.

\begin{figure}[!t] 
    \centering
    \includegraphics[width=0.5\columnwidth,
                     keepaspectratio]{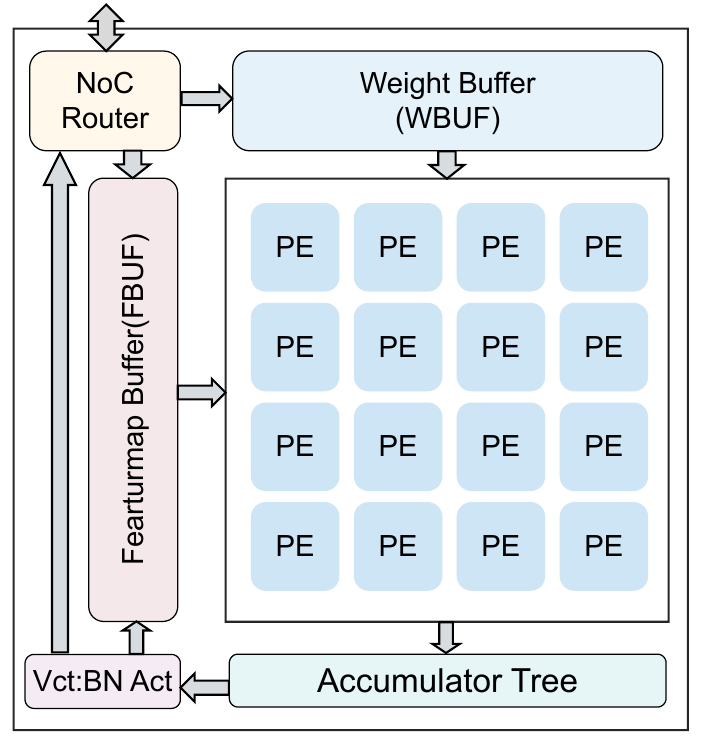}
    \caption {Architecture of a single-engine, \textit{tile}-based accelerator.}
\label{fig:single engine architecture}
\end{figure}

\subsection{Problems of previous works}
In the domain of multi-DNN accelerators, two predominant architectural paradigms have emerged~\cite{CD-MSA}: monolithic PE arrays and multi-subarray (or multi-engine) architectures, with the latter typically implemented using single engine circuits, as illustrated in Fig.~\ref{fig:single engine architecture}.
PREMA exemplifies the former, employing a $128 \times 128$ PE array coupled with twelve on-chip SRAM banks.
However, the monolithic PE-array architecture is ill-suited for the concurrent execution of heterogeneous DNNs.
To overcome this limitation, researchers have proposed Multi-Engine designs.
For example, Planaria partitions a monolithic array into multiple sub-arrays, thereby enabling the concurrent execution of heterogeneous networks.
However, Planaria does not resolve the contention for on-chip SRAM under multi-workloads, which can lead to significant performance degradation.
To address this challenge, MoCA~\cite{moca_cacti} introduces a flexible buffer-allocation and isolation mechanism that effectively mitigates SRAM-sharing conflicts.

In summary, PREMA, Planaria, and MoCA all dedicate hardware resources (a single large PE array or multiple smaller sub-arrays) to execute the same DNN layer, write the intermediate activations to DRAM, and subsequently reload them on-chip for the next layer.
These approaches are essentially instances of LTS.
Consequently, they incur high latency and substantial DRAM-access overhead.

\subsection{Challenges in Supporting TSS-PRM}

To this end, HASP and MARS inspired from single-DNN works~\cite{remap,zhao2025sparsemapsparsetensoraccelerator,fusedlayer,UPot} and proposed the Tile Spatial Schedule (TSS).
As illustrated in Fig.~\ref{fig:DAG to pipeline}, the task mapping of TSS adopts a uniform dataflow across different layers and then decomposes their workloads into smaller units, referred to as \textit{tiles}.
Mapping the individual \textit{tiles} to different engines of the accelerator reduces the number of DRAM accesses to intermediate activations.
Moreover, since a downstream layer can start execution as soon as a single \textit{tile} from its predecessor is generated, it does not need to wait for the completion of the entire preceding layer. Consequently, the overall task latency can be reduced.

The baseline TSS strategy, however, cannot meet the stringent latency requirements of critical tasks in multi-DNN scenarios (\eg, in AR/VR, exceeding 20\,ms for the image-rendering task can induce motion sickness\cite{arvr_20ms}).  
A preemptive variant of Tile Spatial Scheduling (\emph{TSS-PRM}) is therefore indispensable for multi-DNN workloads. However, two fundamental obstacles must be addressed:
\ding{182} The topological complexity of modern DNNs has increased exponentially (Fig.~\ref{fig:Task_scale}). 
For example, a classical network such as ResNet-50 consists of only a few dozen layers, whereas modern transformer models (\eg, Llama-3\cite{Llama-3}) contain more than \(10^{4}\) nodes and \(10^{5}\) edges\cite{deepseek, qwen}.
\ding{183} \emph{TSS-PRM} requires dynamic remapping of DNN graphs onto the network-on-chip (NoC) links, which formulated as a subgraph isomorphism problem—an NP-hard computational challenge~\cite{graph_isomorphism}.


To overcome this challenge, we introduce IsoSched , the first scheduling framework designed to support \emph{TSS-PRM}.
IsoSched  translates the scheduling problem into an integer-linear program (ILP) using a subgraph-isomorphism formulation.

\begin{figure}[!t] 
    \centering
    \includegraphics[width=1.0\columnwidth]{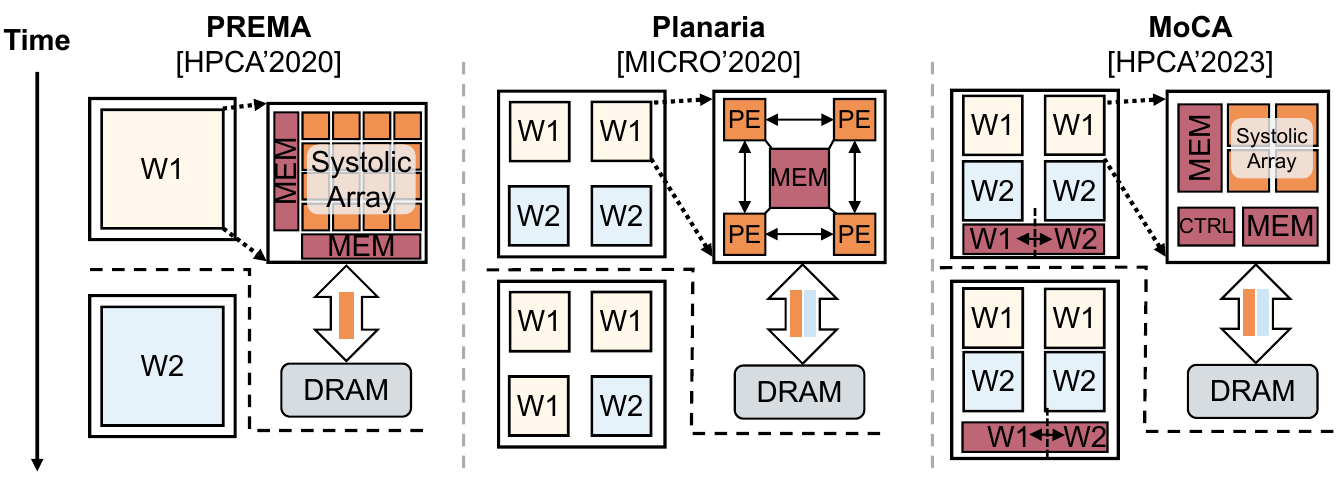}
      \caption{Comparison of the recently \emph{LTS-PRM} works. 
           PREMA supports temporal multi-plexing across different
           DNN workloads. Planaria focuses on dynamic
           partition of the compute resources at the pod granularity, where
           each pod consists of a fixed number of PEs and scratchpads.
           MoCA adaptively partitions both
           the compute and the shared memory resources.}
  \label{fig:multitenancy_comparison}
\end{figure}

\begin{figure}[!t] 
    \centering
    \includegraphics[width=1.0\columnwidth]{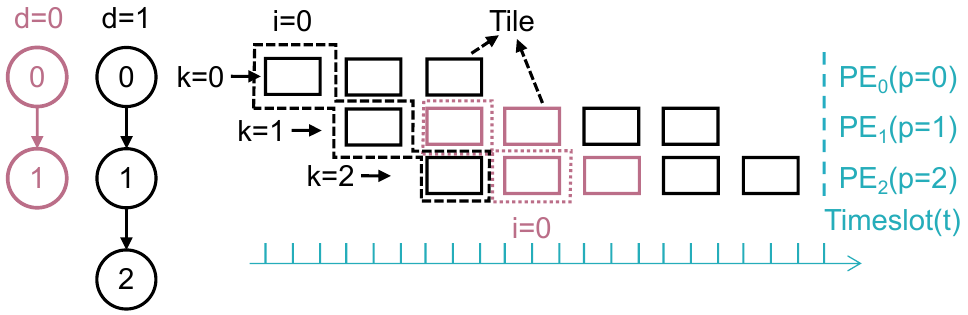}
      \caption{DAG transferred to tile pipeline and allocated to PEs in some \textit{timeslots}.}
  \label{fig:DAG to pipeline}
\end{figure}

\section{IsoSched}
First we present an overview of the IsoSched  workflow (Section~\ref{subsec:isched_overview}).  
Then we formalize the modeling of the multi-DNN scheduling problem (Section~\ref{subsec:problem_formulation}).  
Finally, we introduce the pipeline balancing strategy LCS and the core scheduling algorithm - \textit{MCU} subgraph isomorphism.(Section~\ref{subsec:mcts_ullmann}).
\subsection{IsoSched Workflow Overview}
\label{subsec:isched_overview}

\begin{figure*}[t]
    \centering
    \includegraphics[width=2.0 \columnwidth]{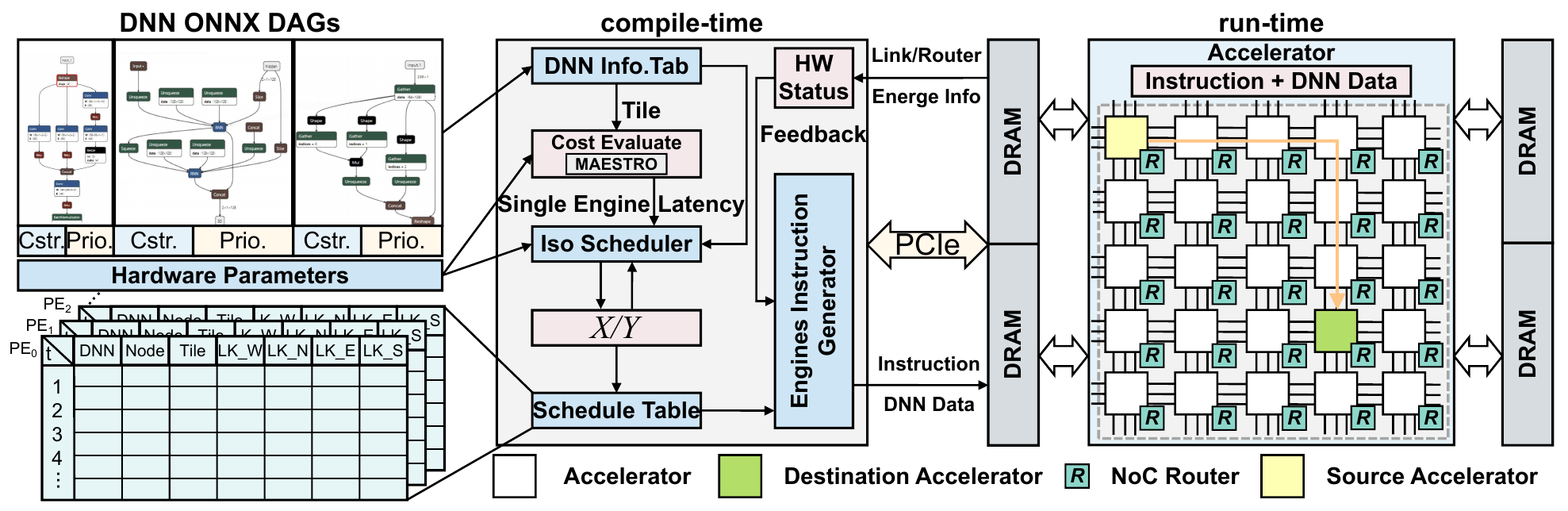}
     \vspace{-3mm}
     \caption{The two stages in the IsoSched  operation process: compile-time and run-time.}
   \label{fig:compile and run time}
\end{figure*}

As shown in Fig.~\ref{fig:compile and run time}, IsoSched  is organised into two phases: compile-time and run-time.

\subsubsection{Compile-time phase}
The compiler accepts a set of DNN DAGs as input, together with their latency constraints (Cstr.) and priority levels (Prio.).
Under a fixed dataflow (weight-stationary for convolutional layers \cite{Eyeriss} and score-stationary for attention layers~\cite{lu2021sanger}), each DAG is partitioned into computation and communication \textit{tiles}.
A cost model provides the single-engine latency of each compute \textit{tile}. In this work, we adopt MAESTRO~\cite{maestro}, whose single-core latency estimation demonstrates a 97\% correlation with silicon measurements.
Iso Scheduler then schedules the compute and communication \textit{tiles} using a graph-matching procedure based on the \textit{MCU} subgraph isomorphism algorithm and updates the state tensors $\mathcal{X}$ (computation) and $\mathcal{Y}$ (communication).
Finally, $\mathcal{X}$ and $\mathcal{Y}$ are translated into a schedule table, from which the instruction streams for each engine and router are generated.

\subsubsection{Run-time phase}
After receiving the instruction streams and \textit{tile} data, the accelerator dispatches them to the designated engines and routers and begins execution. 
During execution, the accelerator records the status of each engine and router and feeds this information back to the host CPU, enabling dynamic regulation of \textit{tile} generation and transmission rates.

\subsubsection{Periodicity}
Following~\cite{CD-MSA,period_survey}, multi-DNN workloads in domains such as AR/VR and autonomous driving are typically executed periodically. 
Accordingly, IsoSched  operates in a cyclic manner, with the scheduling period determined by the scale of hardware resources and the frame rate of upstream sensors, typically ranging from 1\,ms to 1\,s. 
This periodic execution allows the compile-time overhead to be amortized over successive periods, effectively preventing it from becoming a performance bottleneck for the accelerator.


\subsection{Unified Formalisation}
\label{subsec:problem_formulation}

\subsubsection{Key challenges}
Prior studies on multi-DNN execution have not established a rigorous mathematical model for the scheduling problem. In particular, formulating an accurate and fine-grained model specifically tailored to TSS architectures remains a significant challenge~\cite{HASP,MARS}.
Two key issues underpin this difficulty:
\ding{182} \textit{Temporal DAGs}.
    Under TSS, each network is transformed into a temporal DAG.  
    Coordinating multiple such temporal DAGs on a multi-core accelerator is highly nontrivial, as the resulting scheduling tensor spans five dimensions.  
\ding{183} \textit{Joint computation–communication scheduling}.
    The scheduling process must account not only for computational load allocation but also for communication-aware scheduling over the NoC. In particular, each engine in the NoC is typically constrained to four ingress/egress directions (\ie, north, south, east, and west), whereas DNN task DAGs may exhibit fan-in/fan-out degrees greater than four. As a result, NoC links must be split and shared across multiple communication flows. Modeling this complex link-sharing behavior, along with formally expressing bandwidth constraints, presents significant challenges.

\subsubsection{ILP formulation}
The problem is formulated as an ILP. 
Firstly, time is discretized into uniform \textit{timeslots}, with each \textit{timeslot} defined based on the \textit{tile}, as described earlier in Section~\ref{Tile def}.
Secondly, we define two binary tensors:  $\mathcal{X}_{d,i,n,t,p}$, which represents the compute scheduling tensor, and $\mathcal{Y}_{d,i,k,t,\ell}$, which captures the  communication scheduling tensor. Thirdly, the computation must satisfy four constraints: tile compute, tile order, deadline and engine capacity; communication should meet two constraints: link bandwidth constraint and communication cost.
\subsubsection{Scheduling Tensor}
\label{sec:comp_comm_map}

\ding{182} \textit{Compute Scheduling Tensor}.
The scheduling of DNN workloads onto the PE array is expressed by a five‐dimensional tuple $\langle D,I,N,T,P\rangle$,  we define the boolean tensor
\[
  \mathcal{X} \in \{0,1\}^{D\times I\times N\times T\times P}
\]
where we denote \(D\), \(I\), \(N\), \(T\) and \(P\) as the total numbers of DNN tasks, pipeline \textit{tile} groups, DAG nodes, scheduling \textit{timeslots}, and PEs, respectively. Their corresponding indices are denoted as \(d\), \(i\), \(n\), \(t\), and \(p\).
Each element is given by
\begin{equation}
  \mathcal{X}_{d,i,n,t,p} =
  \begin{cases}
    1, & \text{if node } (d,i,n)\text{ is mapped to PE }p\\
       & \text{at timeslot } t\\
    0, & \text{otherwise}
  \end{cases}
\end{equation}

\ding{183} \textit{Communication Scheduling Tensor}

The run-time communication on the on-chip network is described by the
five-dimensional tensor
\[
  \mathcal{Y}\in\{0,1\}^{D\times I\times K\times T\times L}
\]
where \(K\) and \(L\) represent the total numbers of DAG edges and physical links, with corresponding indices \(k\) and \(\ell\).

\begin{equation}
  \mathcal{Y}_{d,i,k,t,\ell} =
  \begin{cases}
    1, & \text{if edge } \delta_{d,k}\text{ is routed through link }\ell\\
       & \text{at timeslot } t\\[4pt]
    0, & \text{otherwise}
  \end{cases}
\end{equation}






\subsubsection{Compute Constraints}

\ding{182} \textit{Tile Compute Constraint}.
A \textit{tile} can be executed only once within its own lifetime:
\begin{equation}
  \sum_{p=0}^{P-1}\sum_{t=S_{d,i,n}}^{L_{d,i,n}}
      \mathcal{X}_{d,i,n,t,p}
    = 1,
  \qquad
  \forall\; \mu_{d,i,n}\in V_{d}
\end{equation}

Here, $V_d$ denotes the complete set of compute tiles for a given DNN task $d$, and $\mu $ represents an individual tile. The variables $S$ and $L$ indicate the start and end \textit{timeslots} of a scheduling period, respectively.


\ding{183} \textit{Tile Order Constraint}.
Execution order must respect data-dependency order:
\begin{equation}
  \begin{gathered}
    \displaystyle
      \sum_{p=0}^{P-1}\sum_{t=S_{d,i,a}}^{L_{d,i,a}}
        \mathcal{X}_{d,i,a,t,p}
      \;-\;
      \sum_{p=0}^{P-1}\sum_{t=S_{d,i,b}}^{L_{d,i,a}}
        \mathcal{X}_{d,i,b,t,p}
      \;\le\;
      -\ell\!\left(\mu_{d,i,a}\right) \\[4pt]
    \forall\,\mu_{d,i,a} \mu_{d,i,b}\in V,\;
      \mu_{d,i,a}\prec\mu_{d,i,b}
  \end{gathered}
\end{equation}

Here, $\ell$ denotes the number of \textit{timeslots} required to execute a given tile.

\ding{184} \textit{Deadline Constraint}.
\begin{equation}
  \begin{gathered}
    \displaystyle
      \sum_{p=0}^{P-1}\sum_{t=S_{d,i,n}}^{L_{d,i,n}}
        \bigl(t + \ell(\mu_{d,i,n})\bigr)\,\mathcal{X}_{d,i,n,t,p}
      - Arr_d < \mathrm{DDL}_d \\[4pt]
    i = I,\; n = N
  \end{gathered}
\end{equation}

\ding{185} \textit{Engine Capacity Constraint}.
At any \textit{timeslot} no more than \(P\) PEs can be occupied (considering the
execution span \(\ell(\mu_{d,i,n})\)):
\begin{equation}
  \sum_{d=0}^{D-1}\sum_{i=0}^{I-1}\sum_{n=0}^{N-1}
  \sum_{r=0}^{\ell(\mu_{d,i,n})-1}\sum_{p=0}^{P-1}
      \mathcal{X}_{d,i,n,t-r,p}
  \;\le\; P
\end{equation}

\subsubsection{Communication Constraints}

\ding{182} \textit{Link-Bandwidth Constraint}.
\begin{equation}
  \begin{gathered}
    \displaystyle
      \sum_{d=0}^{D-1}\!
      \sum_{i=0}^{I-1}\!
      \sum_{k=0}^{K-1}
        \mathcal{Y}_{d,i,k,t,\ell}\,
        \cdot f\!\bigl(\operatorname{bw}(\delta_{d,k}),t,t'\bigr)
      \;\le\; BW\\[4pt]
      \ell\in\{0,\ldots,L-1\},\;
      t\in\bigl \{S_d, \ldots, \,S_d+R \bigr \}
  \end{gathered}
\end{equation}

\begin{equation}
  R \;=\;
  \Bigl\lfloor
    \tfrac{\operatorname{bw}(\delta_{d,k})-1}{BW}
  \Bigr\rfloor
\end{equation}
 $R$ denotes the number of full-bandwidth \textit{timeslots} required to transmit.
\begin{equation}
  t' \;=\;
  idx\!\bigl(\mathcal{Y},d,i,k,t,\,t-R\bigr)
\end{equation}]
$t'$ denotes the starting \textit{timeslot} of a given \textit{tile}.

\begin{equation}
  f(bw,t,t')=
  \begin{cases}
    bw,          & bw \le BW,\; t = t'\\[4pt]
    bw \% BW, & bw > BW,\; t = t' + R\\[4pt]
    BW,          & bw > BW,\; t < t' + R
  \end{cases}
\end{equation}

\ding{183} \textit{Communication Cost}.
For a single dependency edge
\(\mu_{d,i,a}\!\rightarrow\!\mu_{d,i,b}\) we adopt Manhattan distance as communication cost:
\begin{equation}
  C_{\mu_{d,i,a}\prec\mu_{d,i,b}}
  =
  \left|x_a - x_b\right|
  +
  \left|y_a - y_b\right|
\end{equation}
The total communication cost of DAG \(d\) is the sum over all edges:
\begin{equation}
  C_{d}
  =
  \sum_{e=0}^{|\,E_d\,|-1}
    \bigl(
      |x_{e_1} - x_{e_2}|
      +
      |y_{e_1} - y_{e_2}|
    \bigr)
\end{equation}
\subsection{Subgraph-Isomorphism-Based Scheduling}
\label{subsec:mcts_ullmann}

\subsubsection{Balance Pipeline Stage via Layer Concatenate and Split} 
To enable efficient pipeline-level scheduling, we first leverage the DAG-to-Pipeline (D2P) strategy proposed by REMAP~\cite{remap}, which analyzes the computational latency and storage requirements across different dataflows, significantly reducing the exploration space and improving deployment efficiency. However, CNN workloads often exhibit substantial imbalance across layers due to significant variation in their computational demands~\cite{Herald}, posing two key challenges: \ding{182} determining when workload balancing is necessary, and \ding{183} effectively executing balancing strategies with minimal overhead.



To this end, we propose \textbf{L}ayer \textbf{C}oncatenate and \textbf{S}plit (LCS), a workload balancing strategy tailored for CNNs. LCS first determines whether the workload distribution is imbalanced by computing the coefficient of variation (CV), defined as $c_v = \sigma / \mu$, where $\sigma$ and $\mu$ represent the standard deviation and mean of the layer workloads, respectively. Based on empirical evidence, we set a CV threshold of 15\% to trigger balancing, aligning with common definitions of balanced distributions (typically 10\%–20\%)~\cite{CV1,CV2}. We chose to use CV instead of variance since it is dimensionless and facilitates easier comparisons.

Once triggered, LCS evaluates the cost of different layer partitioning and merging strategies across various dimensions, and selects an optimal approach that minimizes latency, buffer usage, and data movement. Unlike conventional node-level DAG splitting, LCS operates at the layer level and incorporates hardware-aware trade-offs to achieve effective workload balance without incurring excessive resource overhead.


\textbf{Layer Concatenate}, as shown in the lower left portion of Fig.~\ref{fig_cat_split}(a), allows us to merge the preceding and succeeding layers into a new \textit{segment} when the computational workload of both layers is relatively small.
We can then map the \textit{segment} onto one engine in the accelerator. While concatenating nodes will increase the required buffer size for each engine. The required buffer size of a \textit{segment} can be defined by the following equations. Specifically, we can estimate the minimal buffer size for a \textit{segment} $s_k$ with multiple layers whose dataflow uses $H$ or $W$ as the outer loop order:
\begin{equation}\label{eq:buffersizey} 
    \resizebox{.9\hsize}{!}{\mbox{BufferSize}($s_k,H$)=$\sum_{l_i\in s_k}(R_i \times W_i \times C_i) + 2\times\max_{l_i\in s_k}(R_i \times S_i \times C_i)$}
\end{equation}
  \begin{equation} \label{eq:buffersizex} 
    \resizebox{.9\hsize}{!}{\mbox{BufferSize}($s_k,W$)=$\sum_{l_i\in s_k}(R_i \times H_i \times C_i) + 2\times\max_{l_i\in s_k}(R_i \times S_i \times C_i)$}
\end{equation}
where the first part is the feature map buffer size with matching outer dimensions of layers, and the second part is the weight buffer size of ping-pong double buffer.

\begin{figure}[!t]
\centering
\includegraphics[width=0.6\columnwidth]{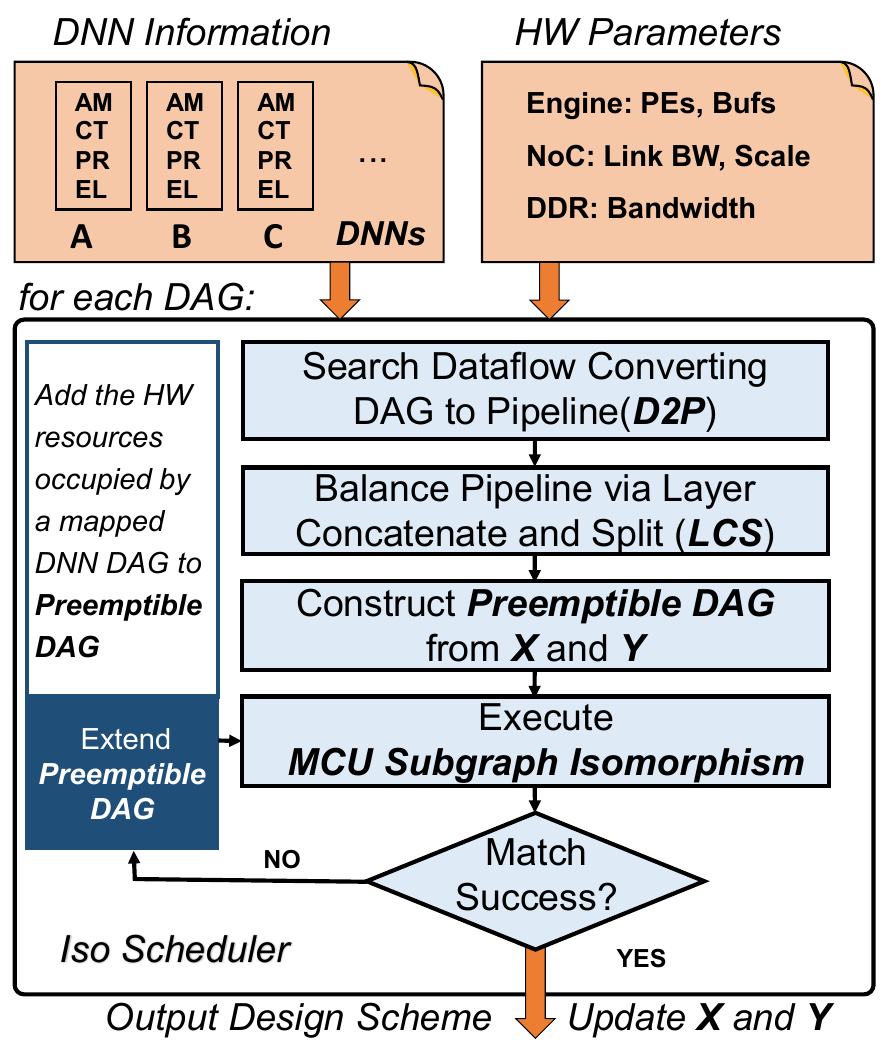}
\caption{The detailed workflow of Iso Scheduler.} 
\label{fig_schedule}
\end{figure}

\begin{figure}[!t]
\centering
\includegraphics[width=1.0\columnwidth]{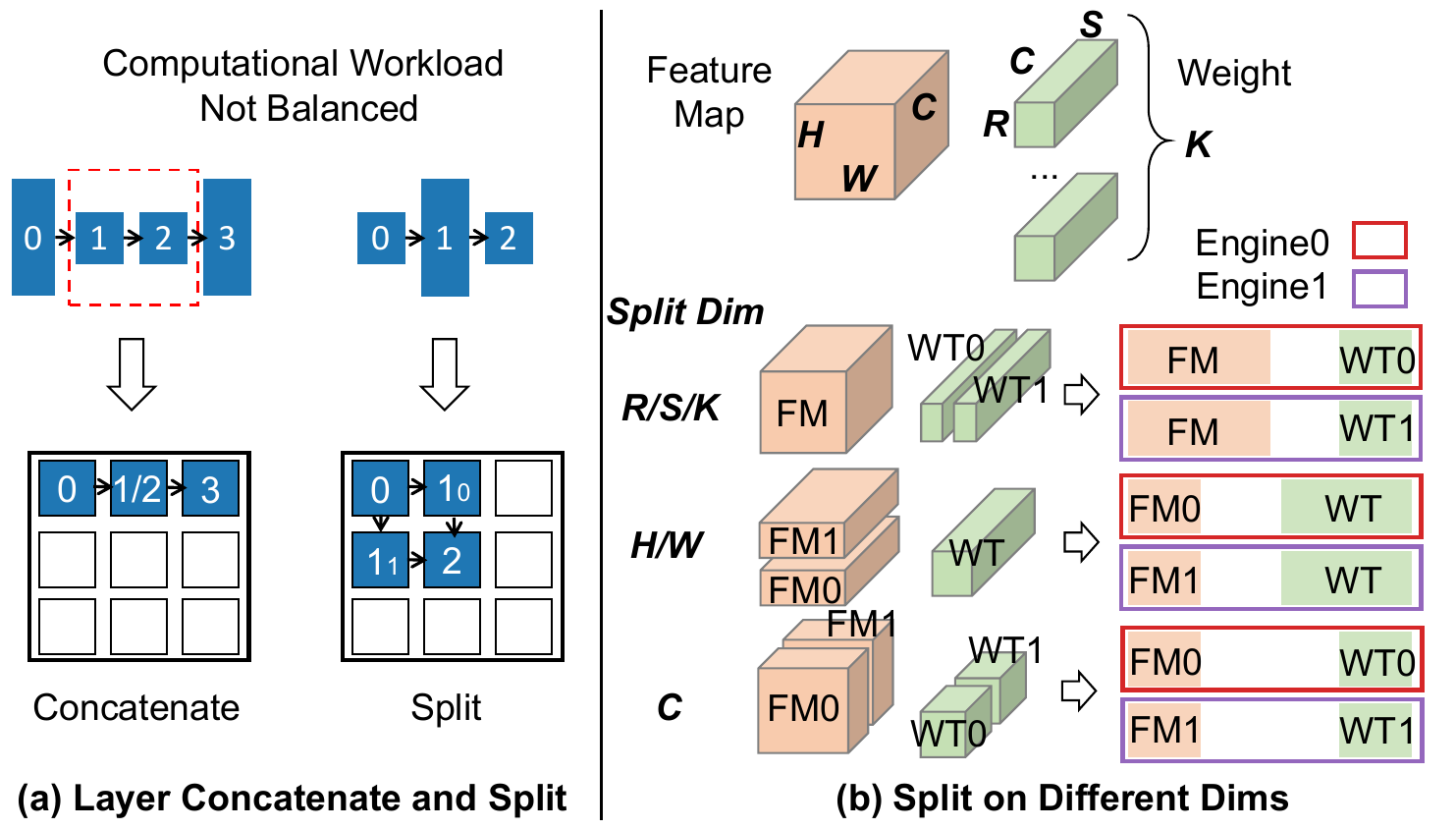}
\caption{(a) Example of layer concatenating and splitting to balance pipeline. (b) Different splitting styles affect data caching and communication among engines.} 
\label{fig_cat_split}
\end{figure}

\textbf{Layer Split} partitions a layer’s workload across engines to improve pipeline balance. The splitting dimension significantly affects buffer usage and communication: splitting along $H/W$ enables each engine to compute full results without partial sum accumulation but requires larger buffer capacity; in contrast, splitting along $C$ reduces memory demand but necessitates additional accumulation. \textsc{LCS} selects the split strategy adaptively based on hardware constraints—preferring $H/W$ when sufficient buffer is available, and $C$ otherwise.

\begin{algorithm}[!t]
\caption{MCU-Based Subgraph Isomorphism}
\label{alg_mcts_main}
\begin{algorithmic}[1]

\Function{MCUSubgraphIsomorphism}{$A, B, T, C$}
  \State \textbf{Input:} Adjacency matrices $A \in \{0,1\}^{n \times n}$ (DNN DAG), $B \in \{0,1\}^{m \times m}$ (preemptible DAG), max iterations $T$, exploration constant $C$
  \State \textbf{Output:} Best mapping $M_{\text{best}}$
  \State root $\gets$ \Call{NewNode}{InitialMapping($n, m$)}
  \State $M_{\text{best}} \gets$ root.$M$, \quad $r_{\text{best}} \gets -\infty$
  \For{$t = 1$ \textbf{to} $T$}
    \State $v \gets$ \Call{Select}{root, $C$}
    \State $u \gets$ \Call{Expand}{$v$}
    \State $r \gets$ \Call{Simulate}{$u, A, B$}
    \State \Call{Backpropagate}{$u, r$}
    \If{$r > r_{\text{best}}$}
      \State $r_{\text{best}} \gets r$;\quad $M_{\text{best}} \gets u.M$
    \EndIf
  \EndFor
  \State \Return $M_{\text{best}}$
\EndFunction

\\

\Function{Select}{$v, C$}
  \While{$v.\textit{children} \neq \varnothing$}
    \State $v \gets \displaystyle \arg\max_{u \in v.\textit{children}} \left( \frac{u.Q}{u.N} + C \sqrt{\frac{\ln v.N}{u.N}} \right)$
  \EndWhile
  \State \Return $v$
\EndFunction

\\

\Function{Expand}{$v$}
  \If{$v$ is terminal}
    \State \Return $v$
  \EndIf
  \State $\mathcal{A} \gets$ \Call{GenerateActions}{$v.M$}
  \State Choose $a \in \mathcal{A}$ uniformly
  \State $M' \gets$ \Call{ApplyAction}{$v.M, a$}
  \State $u \gets$ \Call{NewNode}{$M', v$}
  \State Add $u$ to $v.\textit{children}$
  \State \Return $u$
\EndFunction

\\

\Function{Simulate}{$v, A, B$}
  \State \Return \Call{Evaluate}{$v.M, A, B$}
\EndFunction

\\

\Function{Backpropagate}{$v, r$}
  \While{$v \neq \textbf{null}$}
    \State $v.N \gets v.N + 1$;\quad $v.Q \gets v.Q + r$
    \State $v \gets v.\textit{parent}$
  \EndWhile
\EndFunction

\\

\Function{Evaluate}{$M, A, B$}
  \State $C \gets M^{\mathsf{T}} A M$
  \If{$C \subseteq B$}
    \State \Return $+1$
  \Else
    \State \Return $-1$
  \EndIf
\EndFunction

\\

\Function{GenerateActions}{$M$}
  \State \Return All swaps $(i, j)$
\EndFunction

\end{algorithmic}
\end{algorithm}

\begin{figure*}[t]
    \centering
    \includegraphics[width=2.0\columnwidth]{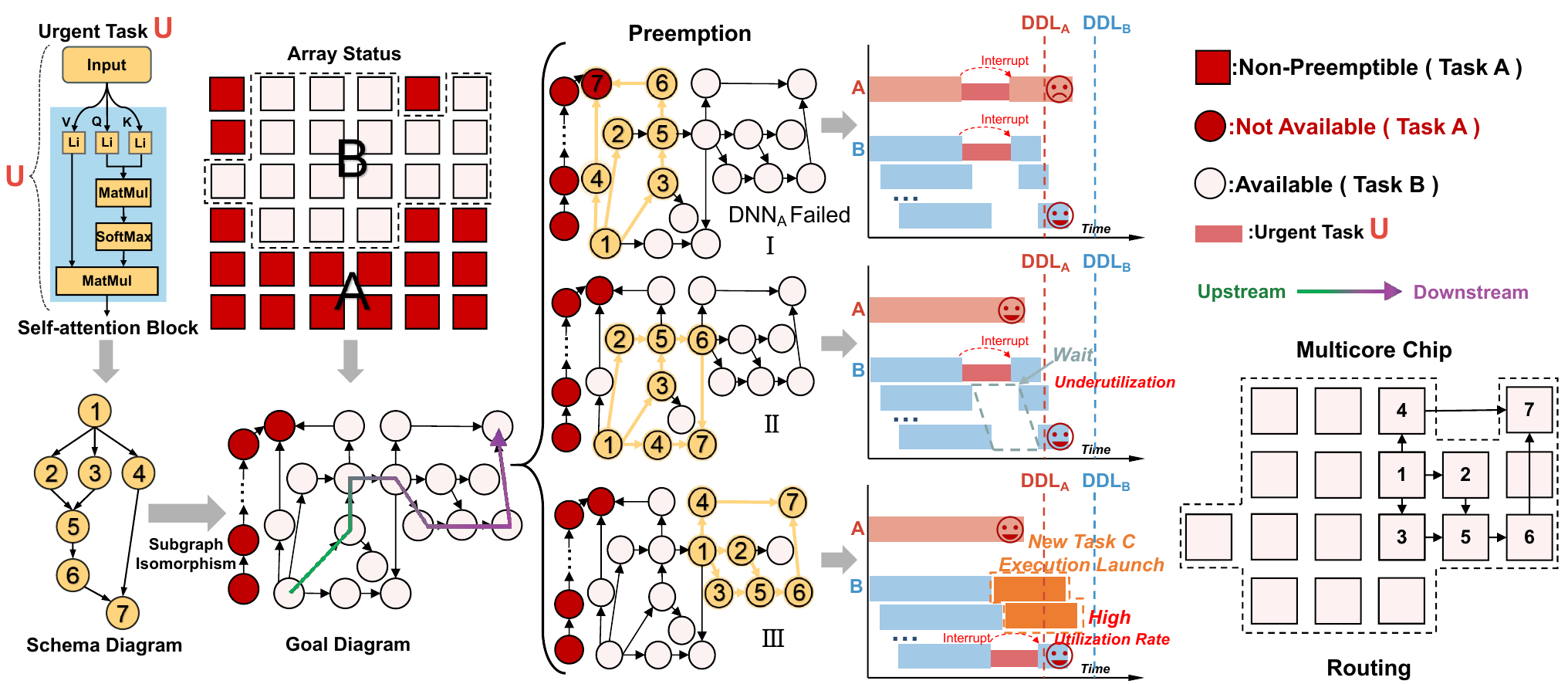}
    \caption{Acceleration performance varies across different subgraph matching schemes.} 
    \label{fig:Subgraph Isomorphism} \vspace{-3mm}
\end{figure*}

\subsubsection{Constructing Preemptible DAG and Executing MCU Subgraph Isomorphism}
To enable preemptive multi-DNN scheduling, we construct a preemptible DAG based on the current computation and communication scheduling matrices, $\mathcal{X}$ and $\mathcal{Y}$. This DAG is dynamically updated by incorporating hardware resources occupied by previously scheduled DNNs. The system then invokes the \textit{MCU} subgraph isomorphism to determine whether the topology of a newly arriving DNN DAG can be matched onto the preemptible DAG. If a match is successful, the corresponding scheduling plan is generated, including PE assignments and NoC routing configurations, while the scheduling matrices $\mathcal{X}$ and $\mathcal{Y}$ are updated accordingly. Otherwise, additional scheduled tasks are selectively integrated into the preemptible DAG based on their \textit{latency slack}, and the matching process is retried.
The selection is guided by a \textit{latency slack}:
\begin{equation}
  W_d
  \;=\;
  \frac{t^{\mathrm{ddl}}_d - t^{\mathrm{now}}}{\tau_d}
  \,\bigg/\,
  \frac{P_d}{\displaystyle\sum_j P_j}
\end{equation}
where  
\( \tau_d \) represents the remaining execution time of DNN~\( d \),  
and \( P_d \) is the priority level of DNN~\( d \).  
A larger \( P_d \) indicates higher execution urgency.

When multiple matching results exist, the IsoSched scheduler selects the option that minimizes disruption to ongoing tasks. For example, as illustrated in Fig.~\ref{fig:Subgraph Isomorphism}, consider an urgent DNN task~U to be scheduled. Based on the current deployment status of DNN tasks on hardware—retrieved from the scheduling matrices $\mathcal{X}$ and $\mathcal{Y}$. After performing \textit{MCU} subgraph isomorphism, three candidate preemption schemes are identified:
\ding{182} Scheme~I preempts critical DNN task~A, resulting in its execution failure of task A. Therefore, this scheme is considered unacceptable.
\ding{183} Scheme~II preempts only non-critical task~B. However, it occupies B's upstream pipeline engines, causing downstream engines of task B to idle and significantly degrading throughput.
\ding{184} Scheme~III preempts B's downstream engines, minimally impacting B. Most upstream engines remain active, allowing subsequent tasks to execute immediately after  upstream engines of task B completes. As a result, this scheme offers the highest  throughput.
IsoSched  thus selects Scheme~III as the final mapping for task~U.

Formally, algorithm~\ref{alg_mcts_main} presents the MCTS-based subgraph isomorphism procedure. Given a task DAG $A \in \{0,1\}^{n \times n}$ and a preemptible DAG $B \in \{0,1\}^{m \times m}$, the algorithm begins by constructing the root node using a randomly initialized mapping matrix $M$ (Lines~2--5), and initializes the best reward $r_{\text{best}}$ and best mapping $M_{\text{best}}$. It then proceeds with $T$ rounds of iterative search (Line~6). In each iteration, the algorithm performs four standard MCTS steps: \textsc{Select}, \textsc{Expand}, \textsc{Simulate}, and \textsc{Backpropagate}. During \textsc{Select} (Lines~15--18), a node is recursively selected from the search tree by applying the Upper Confidence Bound (UCB) formula to balance exploitation and exploration, i.e., selecting the child $u$ that maximizes $\frac{u.Q}{u.N} + C \sqrt{\frac{\ln v.N}{u.N}}$. The selected node is then expanded (Lines~20--28) by first checking whether it is terminal; if not, the algorithm enumerates all possible swap actions $\mathcal{A}$ on the current mapping $v.M$ via the \textsc{GenerateActions} function (Lines~45--46), samples one action $a \in \mathcal{A}$ uniformly, applies it to generate a new mapping $M'$, and adds the corresponding child node $u$ to the tree. In the \textsc{Simulate} phase (Lines~30--31), the algorithm calls the \textsc{Evaluate} function (Lines~38--43), which computes $C = M^{\mathsf{T}} A M$ and checks whether $C \subseteq B$ to determine subgraph containment, returning a reward of $+1$ for success and $-1$ otherwise. The reward is then propagated back to the root during \textsc{Backpropagate} (Lines~33--36), where the visit count and accumulated reward statistics of each node along the path are updated. If the current reward exceeds the best reward found so far, the best score and corresponding mapping are updated accordingly (Lines~11--12). After completing $T$ iterations, the algorithm returns the best mapping $M_{\text{best}}$ (Line~13). All matrices $M$, $A$, $B$, and $C$ are represented and processed in the CSR format.

\subsubsection{Preemption overhead} Prior to preemption, the intermediate data of the preempted task is offloaded to DRAM via newly assigned communication links, while the weights of the incoming task are transmitted through reconfiguration links to overwrite the original weights. After preemption, the original weights are reloaded via the reconfiguration links, and the communication paths of the preempted task are restored.

DRAM link operations are confined to compile-time, whereas weight transfers impact both compile-time and run-time. The associated latency is modeled as $SIZEOF(WT)/BW$.

\begin{table}[t]
\caption{Hardware resource constrains of different platforms}
\centering
\resizebox{0.5\textwidth}{!}{ 
\begin{tabular}{|l|c|c|c|}
\hline
\textbf{Platform} & \textbf{MACs in Engine} & \textbf{Number of Engine} & \textbf{Clock Frequency} \\
\hline
Edge & 64 & $128\times $128  & 700MHz\\
\hline
Cloud & 128 & $128\times $128  & 700MHz\\
\hline
\end{tabular}
}

\label{tab_platforms}

\end{table}

\section{Evaluation}

\subsection{Methodology}
\subsubsection{Hardware Modeling} 
Following Planaria\cite{planaria} and MoCA\cite{moca_cacti}, we implement the proposed engines and NoC in Verilog and synthesize them using Synopsys Design Compiler (T-2022.03-SP5) with the FreePDK 45nm~\cite{FreePDK45} standard cell library to extract their power and area.
We model the on-chip SRAM using CACTI-P \cite{moca_cacti} that provides both energy and area.  
The NoC is modeled with McPAT 1.3 \cite{McPAT}; the per-hop energy cost is estimated to be 0.64 pJ/bit.  
Platform description in Table~\ref{tab_platforms}.

\subsubsection{Simulation Setup}
We develop a cycle-accurate simulator that provides the cycle counts and statistics for energy measurements for each DNN using the modeling described above. 
We include all the overheads of reconfiguration, instruction fetch, off-chip memory accesses, etc. 
We verify the cycle counts with our Verilog implementations.


\subsubsection{Workloads}
To evaluate how IsoSched  performs under DNN DAGs of varying complexity, we construct three multi-DNN workloads: Simple, Middle, and Complex.
As shown in Fig.~\ref{fig:speedup}, the Simple workload, following Herald \cite{HDA}, targets AR/VR scenarios and consists of MobileNetV2 \cite{Mobilenetv2}, ResNet-50 \cite{resnet50}, and EfficientNet \cite{Efficientnet}.  
The Middle workload is derived from AutoDAG \cite{atomic_DAG} and represents Neural Architecture Search (NAS) scenarios with UNet \cite{U-net}, NASNet \cite{NASNet}, and PNASNet \cite{PNASNet}.  
For the Complex workload, we examine the distribution of edges and nodes in contemporary DNNs and select three networks that contain more than 5{,}000 nodes and more than 10{,}000 edges—namely Deepseek-7B\cite{deepseek}, Qwen-7B\cite{qwen}, and Llama-3-8B\cite{Llama-3}.
\begin{figure}[t]
    \centering
    \includegraphics[width=1.0\columnwidth]{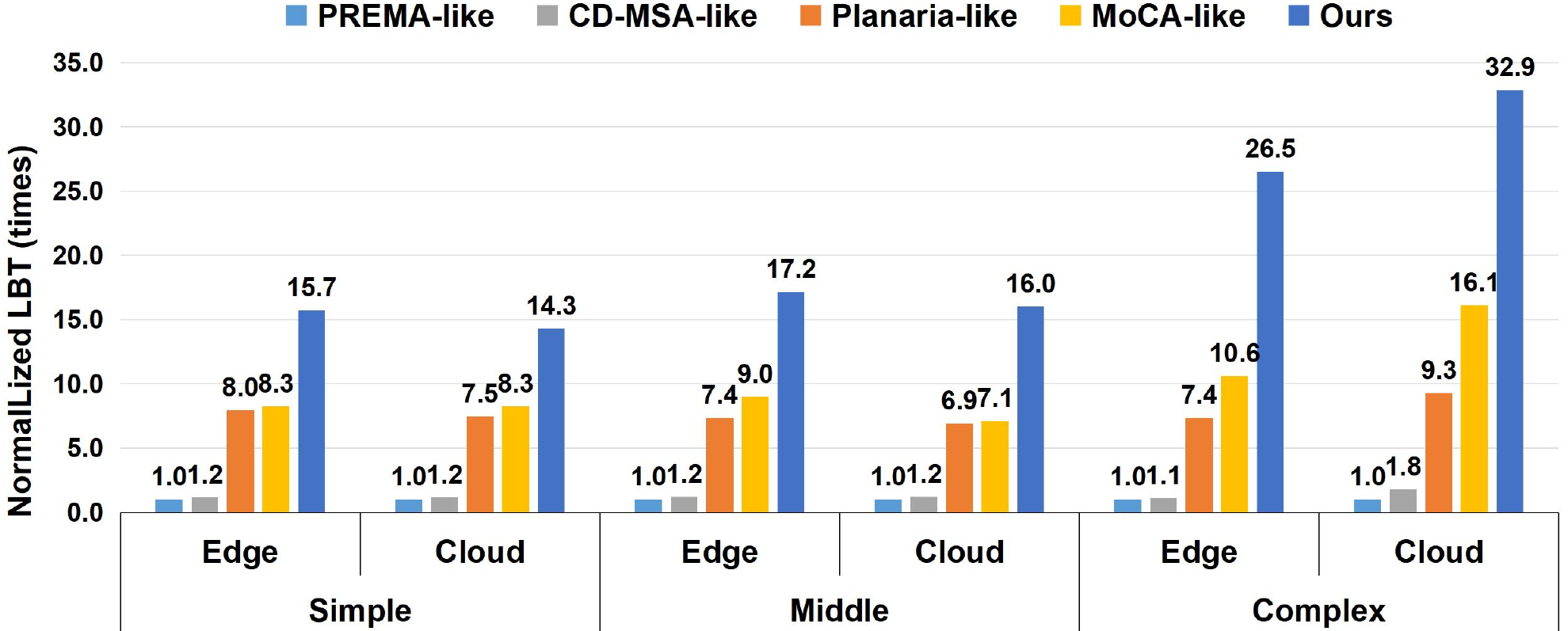}
    \caption{Latency-Bound Throughput (LBT) performance (normalized) across various platforms and workloads of different complexities.} 
    \label{fig:LBT} \vspace{-3mm}
\end{figure}
\subsubsection{Metric}  
We evaluate the proposed method using four metrics. Specifically,
\ding{182} Service-Level Agreement (SLA): This metric, following Planaria, measures the rate at which task latency requirements are satisfied.
\ding{183} Latency-Bound Throughput (LBT): Following PREMA \cite{PREMA}, Planaria, and CD-MSA\cite{CD-MSA}, this metric is defined as the maximum queries-per-second ($1/\lambda$) achieved by the system under a Poisson arrival rate~$\lambda$. 
According to MLPerf \cite{MLPerf}, meeting the SLA means completing an image-classification or object-detection task 99\% of the time, and a translation task 97\% of the time. 
\ding{184} Speedup reflects the reduction in task latency compared to the baseline.
\ding{185} Energy Efficiency denotes the amount of throughput achieved per unit of energy consumption~\cite{xia2022energy}.

\begin{figure*}[t]
    \centering
    \includegraphics[width=2.0\columnwidth]{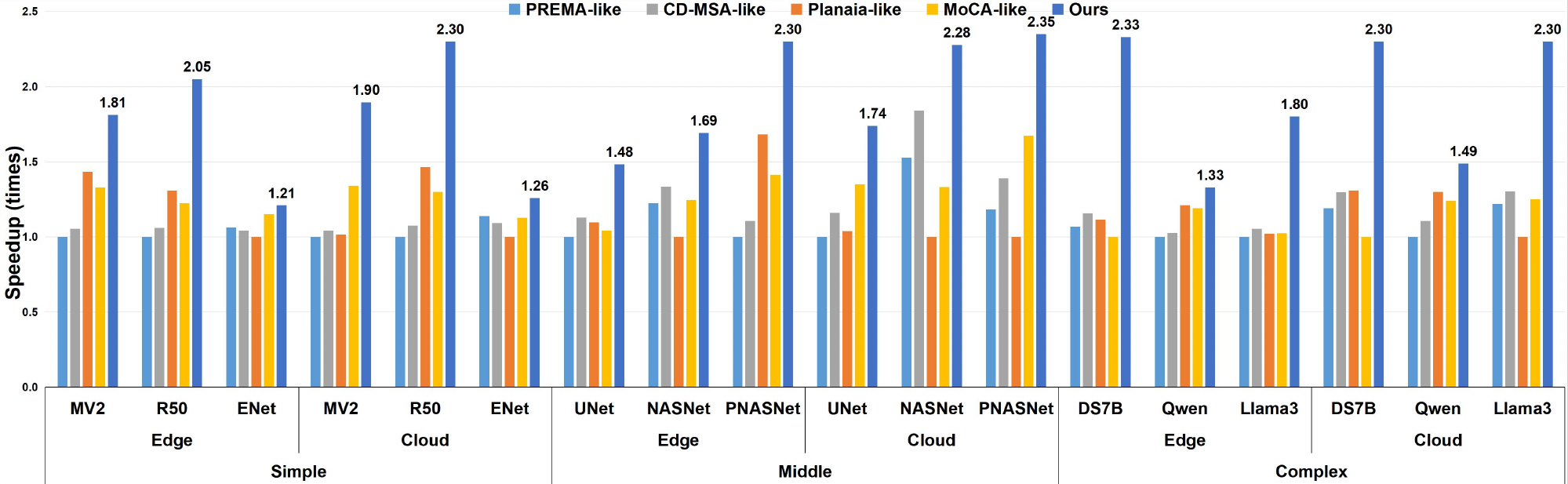}
    \caption{Speedup performance of various neural networks across different models under various platforms.} 
    \label{fig:speedup} \vspace{-3mm}
\end{figure*}

\subsubsection{Baselines}
IsoSched adopts the \emph{TSS-PRM} technique.  
To quantify its effectiveness we compare it against two baseline groups, \emph{LTS-PRM} and \emph{TSS-NPRM}. Specifically, for \emph{LTS-PRM}, We compare PREMA, CD-MSA, Planaria  MoCA and our work under LBT, Speedup and Energy; for \emph{TSS-NPRM}, We use HASP\cite{HASP} as the sole representative and evaluate LBT and SLA Stat. Rate to determine whether urgent tasks still meet their latency bounds under multi-DNN resource contention.
Only HASP is selected because \ding{182} few studies explore \emph{TSS-NPRM}, and \ding{183} MARS focuses on PCB-level communication scheduling, which is orthogonal to our problem domain.  
\emph{LTS-NPRM} is excluded for the same reason: its methodology and application scenario differ markedly from those of IsoSched.


Notably, to ensure a fair comparison, all baseline frameworks are configured with an equal number of MAC by modifying their corresponding Verilog implementations (\ie, $64 \times 128 \times 128$ and $128 \times 128 \times 128$ for Edge and Cloud, respectively).
For PREMA, Planaria and MoCA, we adopt their official open-source releases PREMA\footnotemark[1], Planaria\footnotemark[2], MoCA\footnotemark[3] and adjust MAC configurations named PREMA-like, Planaria-like and MoCA-like. 
Since the source code for CD-MSA and HASP is unavailable, we re-implement them based on the descriptions provided in their respective publications, and refer to the resulting versions as CD-MSA-like and HASP-like.

\footnotetext[1]{PREMA: \url{https://github.com/agongee/prema_sim}}
\footnotetext[2]{Planaria: \url{https://github.com/he-actlab/planaria.code}}
\footnotetext[3]{MoCA: \url{https://github.com/ucb-bar/MoCA}}


\subsection{Comparison with LTS--PRM Baselines}
\label{subsec:lts_prm_results}

\subsubsection{Latency-Bound Throughput}

As shown in Fig.~\ref{fig:LBT}, across tasks with different topological complexities (Simple, Middle, Complex), compared to \emph{LTS-PRM} approaches including PREMA-like, Planaria-like, CD-MSA-like, and MoCA-like, IsoSched achieves average improvements in latency-bound throughput by $\times$20.4, $\times$2.6, $\times$15.8, and $\times$2.1, respectively. 
In addition, we observe that our work exhibits progressively better performance as the topological complexity increases. Specifically, compared to PREMA-like, IsoSched achieves average improvements of $\times$15.0, $\times$16.6, and $\times$29.7 on Simple, Middle and Complex topologies, respectively.
This improvement is primarily attributed to the algorithm's ability to effectively model and exploit the structural characteristics of complex topologies during subgraph matching.
Specifically, a key contributor to this performance gain is the Ullmann-based subgraph isomorphism algorithm accelerated by MCTS. By guiding the search with statistical sampling, MCTS efficiently explores the large and discrete mapping space and quickly identifies topology-consistent subgraph matches. These matches significantly reduce disruption to resident tasks—thereby enabling high utilization under latency constraints.

\subsubsection{Speedup}
As shown in Fig.~\ref{fig:speedup}, compared to \emph{LTS-PRM} approaches (\ie, PREMA-like, Planaria-like, CD-MSA-like and MoCA-like frameworks), IsoSched achieves average speedup improvements of $\times$1.9, $\times$1.6, $\times$1.6, and $\times$1.5, respectively. These improvements are primarily attributed to the TSS mechanism, which pipeline \textit{tiles} over on-chip links instead of waiting for full-layer outputs to be written to and read from off-chip DRAM, as is typical in traditional LTS frameworks. This tile pipeline allows downstream layers to begin execution as soon as partial results from upstream layers become available, effectively enabling inter-layer execution overlap and significantly reducing total latency. 
Moreover, IsoSched further shortens the overall completion time by incorporating preemptive scheduling through Ullmann-based subgraph isomorphism.

\subsubsection{Energy Efficiency}
\begin{figure}[t]
    \centering
    \includegraphics[width=1.0\columnwidth]{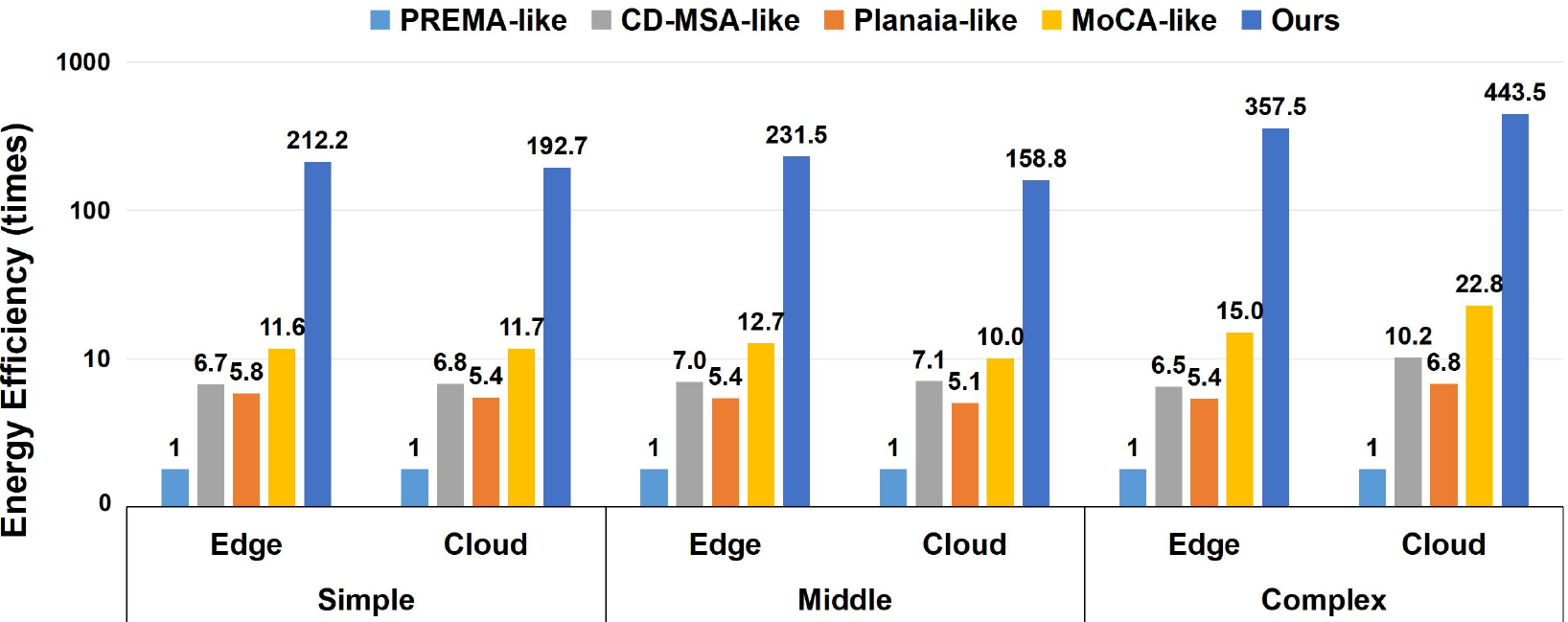}
    \caption{Energy efficiency performance (normalized) across various platforms and workloads of different complexities.} 
    \label{fig:energy} \vspace{-3mm}
\end{figure}

As shown in Fig.~\ref{fig:energy}, in terms of energy efficiency, compared to \emph{LTS-PRM} approaches including PREMA-like, Planaria-like, CD-MSA-like, and MoCA-like, IsoSched achieves average improvements in energy efficiency by $\times$266.0, $\times$46.3, $\times$35.7, and $\times$18.7, respectively. These gains are attributed to several key factors. Firstly, IsoSched reduces communication energy by replacing off-chip DRAM accesses with on-chip data transfers, effectively utilizing on-chip links. Secondly, the use of MCU subgraph isomorphism shortens scheduling latency and enhances hardware resource utilization. Thirdly, LCS improves pipeline efficiency, further enhancing overall energy efficiency.



\begin{figure}[t]
    \centering
    \includegraphics[width=1.0\columnwidth]{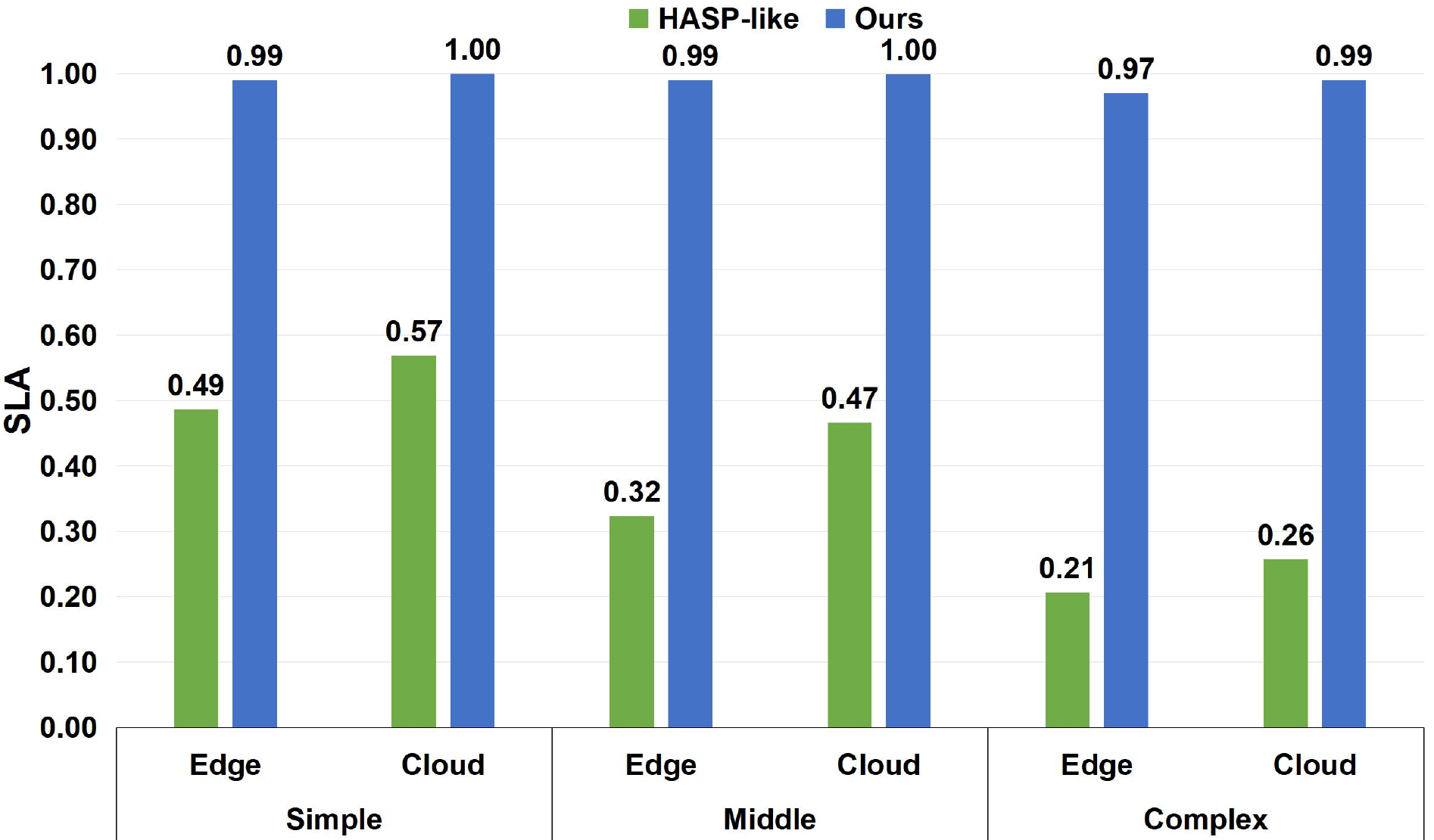}
    \caption{The results of comparative experiments of SLA performance between HASP and our work under various workloads.} 
    \label{fig:sla} \vspace{-3mm}
\end{figure}
\begin{figure}[t]
    \centering
    \includegraphics[width=0.9\columnwidth]{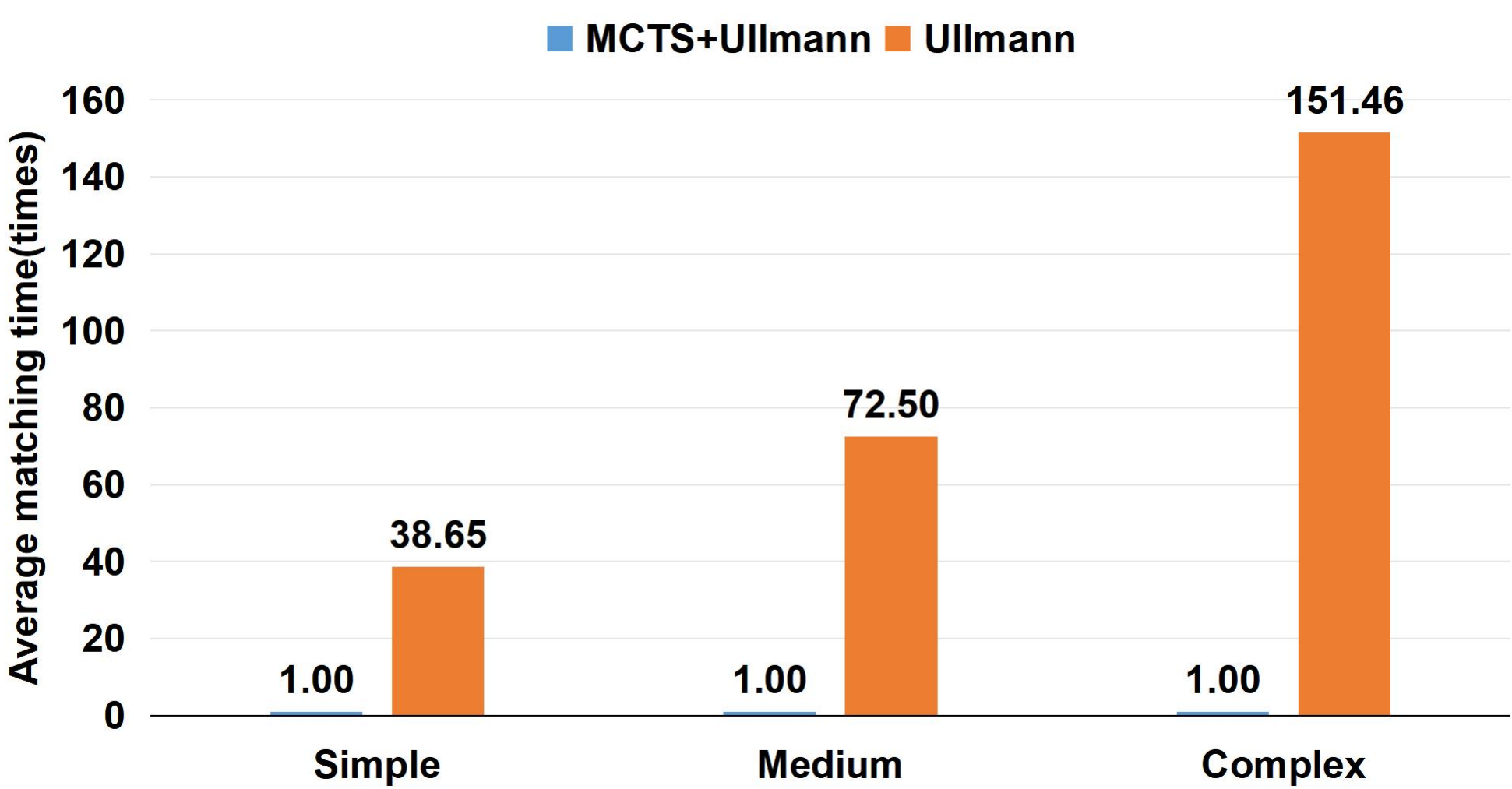}
    \caption{Ratio of average matching time for Ullmann algorithm with MCTS enhancement versus without MCTS across various workloads.} 
    \label{fig:ablation_MCTS} \vspace{-3mm}
\end{figure}
\begin{figure}[t]
    \centering
    \includegraphics[width=1.0\columnwidth]{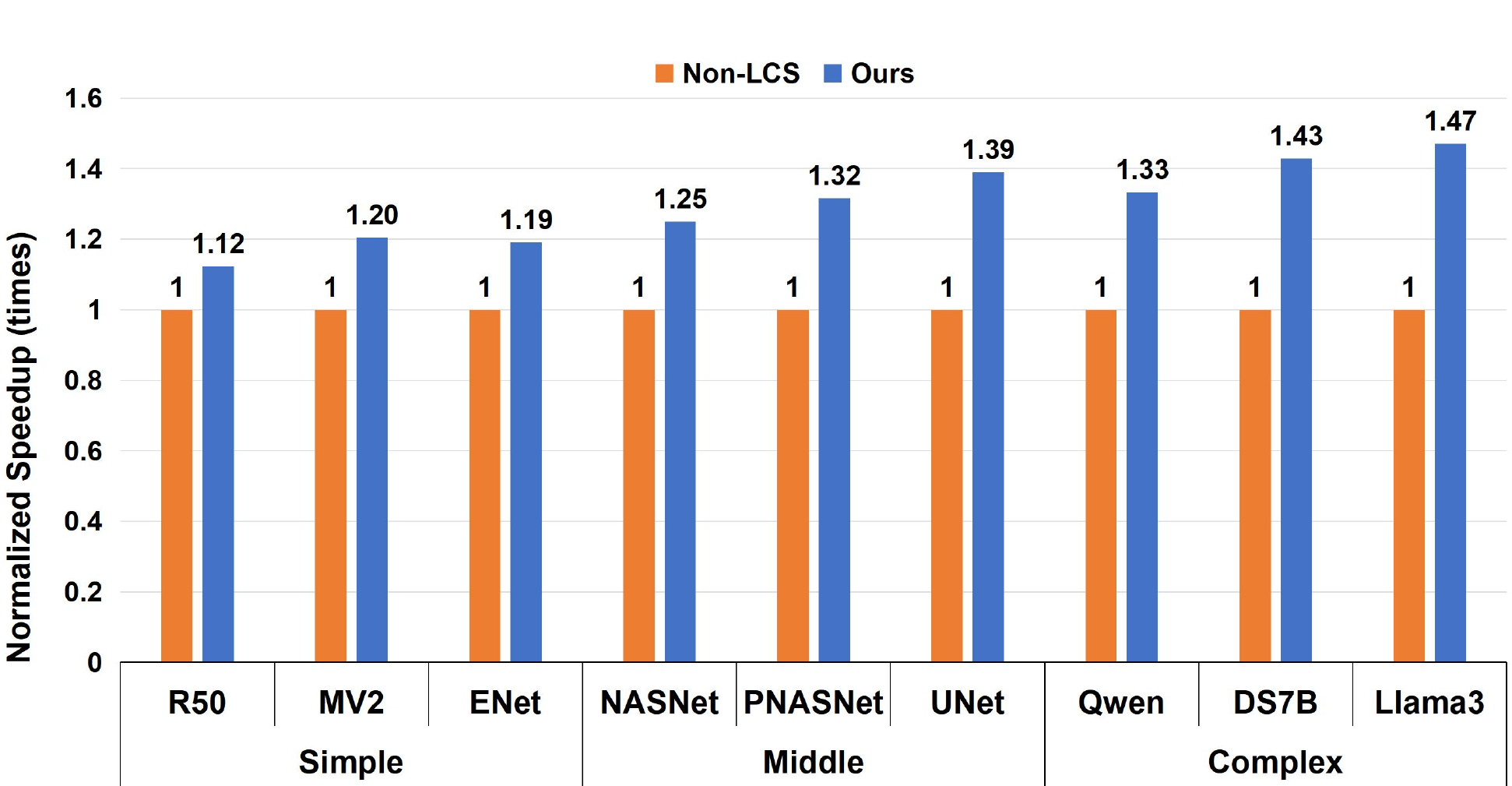}
    \caption{Speedup ratio comparison of with LCS versus without LCS configurations across various neural networks on Cloud platform.} 
    \label{fig:ablation_LCS} \vspace{-3mm}
\end{figure}
\begin{figure}[t]
    \centering
    \includegraphics[width=1.0\columnwidth]{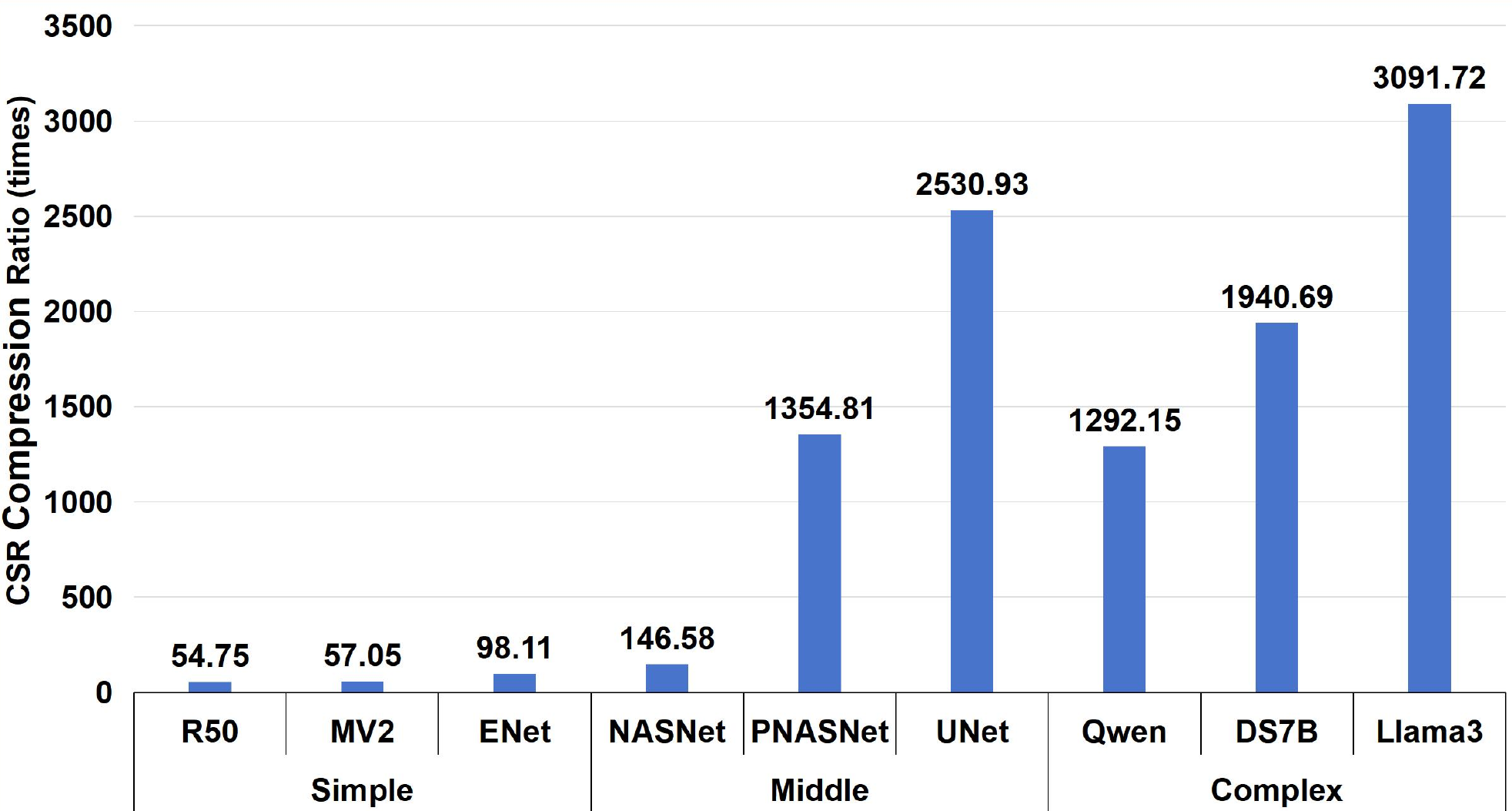}
    \caption{Compression ratio achieved by CSR versus baseline on Cloud platform.} 
    \label{fig:ablation_CSR} \vspace{-3mm}
\end{figure}

\subsection{Comparison with TSS-NPRM Baseline}
\label{subsec:tss_nprm_results}
As shown in Fig.~\ref{fig:sla}, compared with the \emph{TSS-NPRM} baseline (\ie, HASP-like), the SLA satisfaction rate is improved by $\times$1.9, $\times$2.6, and $\times$4.3 under the Simple, Middle, and Complex settings, respectively. The lower SLA satisfaction of HASP stems from its NPRM scheduling policy, which prevents it from prioritizing critical tasks when resource contention arises. In contrast, IsoSched enables PRM scheduling through MCU subgraph isomorphism, allowing urgent tasks to be served earlier. Additionally, LCS balances the tile pipeline, stabilizing stage latencies and sustaining PE utilization to improve deadline satisfaction.

\subsection{Ablation Study: Effectiveness of MCTS, LCS and CSR}
To demonstrate the effectiveness of MCTS, LCS and CSR, we conduct ablation experiments. 
As shown in Fig.~\ref{fig:ablation_MCTS}, MCTS achieves average matching-time reductions of $\times$38.7, $\times$72.5, and $\times$151.5; as shown in Fig.~\ref{fig:ablation_LCS}, LCS yields normalized speedups of $\times$1.2, $\times$1.3, and $\times$1.4, respectively; and as shown in Fig.~\ref{fig:ablation_CSR}, CSR attains average compression ratios of $\times$70.0, $\times$1{,}344.1, and $\times$2{,}108.2 under the Simple, Middle, and Complex settings. These results demonstrate that MCTS effectively reduces the search overhead of the Ullmann algorithm; LCS balances the pipeline to improve hardware utilization and speedup while reducing latency; and CSR significantly reduces the memory footprint of the matching matrix during Ullmann-based subgraph isomorphism.


\section{Conclusion}
\label{sec:conclusion}

The increasing demand to concurrently run multiple DNNs on shared accelerators exposes a key limitation of Layer Temporal Scheduling (LTS), where intermediate activations are staged in DRAM, incurring significant latency and energy costs. Tile Spatial Scheduling (TSS) mitigates these overheads by streaming \textit{tiles} via on-chip links and enabling early execution of downstream layers, but prior TSS solutions lack support for preemption. This paper introduces IsoSched, the first preemptive scheduling framework for multi-DNN execution under TSS. IsoSched jointly optimizes compute and memory mappings through an ILP formulation with subgraph isomorphism, builds a preemptible DAG, and uses MCTS-accelerated Ullmann matching to identify minimally disruptive remaps. It further reduces memory and matching overhead via CSR encoding and balances pipeline depth through Layer Concatenate and Split (LCS). Experiments on diverse topologies demonstrate that IsoSched significantly improves latency-bound throughput, speedup, and energy efficiency over \emph{LTS-PRM} baselines, while achieving higher critical task satisfaction compared to the \emph{TSS-NPRM} baseline.

\bibliography{IEEEabrv,refs}

\begin{thebibliography}{10}
\providecommand{\url}[1]{#1}
\csname url@samestyle\endcsname
\providecommand{\newblock}{\relax}
\providecommand{\bibinfo}[2]{#2}
\providecommand{\BIBentrySTDinterwordspacing}{\spaceskip=0pt\relax}
\providecommand{\BIBentryALTinterwordstretchfactor}{4}
\providecommand{\BIBentryALTinterwordspacing}{\spaceskip=\fontdimen2\font plus
\BIBentryALTinterwordstretchfactor\fontdimen3\font minus \fontdimen4\font\relax}
\providecommand{\BIBforeignlanguage}[2]{{%
\expandafter\ifx\csname l@#1\endcsname\relax
\typeout{** WARNING: IEEEtran.bst: No hyphenation pattern has been}%
\typeout{** loaded for the language `#1'. Using the pattern for}%
\typeout{** the default language instead.}%
\else
\language=\csname l@#1\endcsname
\fi
#2}}
\providecommand{\BIBdecl}{\relax}
\BIBdecl

\bibitem{GraphRef1}
J.~Zhou, G.~Cui, S.~Hu, Z.~Zhang, C.~Yang, Z.~Liu, L.~Wang, C.~Li, and M.~Sun, ``Graph neural networks: A review of methods and applications,'' \emph{AI open}, vol.~1, pp. 57--81, 2020.

\bibitem{GraphRef2}
F.~Wu, A.~Souza, T.~Zhang, C.~Fifty, T.~Yu, and K.~Weinberger, ``Simplifying graph convolutional networks,'' in \emph{International conference on machine learning}.\hskip 1em plus 0.5em minus 0.4em\relax Pmlr, 2019, pp. 6861--6871.

\bibitem{GraphRef3}
F.~Gama, A.~G. Marques, G.~Leus, and A.~Ribeiro, ``Convolutional neural network architectures for signals supported on graphs,'' \emph{IEEE Transactions on Signal Processing}, vol.~67, no.~4, pp. 1034--1049, 2018.

\bibitem{GraphRef4}
S.~Abadal, A.~Jain, R.~Guirado, J.~L{\'o}pez-Alonso, and E.~Alarc{\'o}n, ``Computing graph neural networks: A survey from algorithms to accelerators,'' \emph{ACM Computing Surveys (CSUR)}, vol.~54, no.~9, pp. 1--38, 2021.

\bibitem{GraphRef5}
L.~Ruiz, F.~Gama, and A.~Ribeiro, ``Gated graph recurrent neural networks,'' \emph{IEEE Transactions on Signal Processing}, vol.~68, pp. 6303--6318, 2020.

\bibitem{zhao2025nms}
B.~Zhao, H.~Huang, Q.~Dang, W.~Zhao, T.~Xia, and P.~Ren, ``Nms: Efficient edge dnn training via near-memory sampling on manifolds,'' \emph{arXiv preprint arXiv:2508.02313}, 2025.

\bibitem{Eyeriss}
Y.-H. Chen, T.~Krishna, J.~S. Emer, and V.~Sze, ``Eyeriss: An energy-efficient reconfigurable accelerator for deep convolutional neural networks,'' \emph{IEEE journal of solid-state circuits}, vol.~52, no.~1, pp. 127--138, 2016.

\bibitem{NVDLA}
``Nvdla deep learning accelerator,'' \url{http://nvdla.org}, 2017.

\bibitem{ShiDianNao}
Z.~Du, R.~Fasthuber, T.~Chen, P.~Ienne, L.~Li, T.~Luo, X.~Feng, Y.~Chen, and O.~Temam, ``Shidiannao: Shifting vision processing closer to the sensor,'' in \emph{Proceedings of the 42nd annual international symposium on computer architecture}, 2015, pp. 92--104.

\bibitem{fusedlayer}
M.~Alwani, H.~Chen, M.~Ferdman, and P.~Milder, ``Fused-layer cnn accelerators,'' in \emph{2016 49th Annual IEEE/ACM International Symposium on Microarchitecture (MICRO)}.\hskip 1em plus 0.5em minus 0.4em\relax IEEE, 2016, pp. 1--12.

\bibitem{remap}
B.~Zhao, T.~Xia, H.~Zhai, F.~Ma, Y.~Du, H.~Chang, W.~Zhao, and P.~Ren, ``Remap: A spatiotemporal cnn accelerator optimization methodology and toolkit thereof,'' \emph{IEEE Transactions on Computer-Aided Design of Integrated Circuits and Systems}, 2022.

\bibitem{isos}
Y.~Yang, J.~S. Emer, and D.~Sanchez, ``Isosceles: Accelerating sparse cnns through inter-layer pipelining,'' in \emph{2023 IEEE International Symposium on High-Performance Computer Architecture (HPCA)}.\hskip 1em plus 0.5em minus 0.4em\relax IEEE, 2023, pp. 598--610.

\bibitem{maestro}
H.~Kwon, P.~Chatarasi, V.~Sarkar, T.~Krishna, M.~Pellauer, and A.~Parashar, ``Maestro: A data-centric approach to understand reuse, performance, and hardware cost of dnn mappings,'' \emph{IEEE micro}, vol.~40, no.~3, pp. 20--29, 2020.

\bibitem{autocar_re0}
H.~Fujiyoshi, T.~Hirakawa, and T.~Yamashita, ``Deep learning-based image recognition for autonomous driving,'' \emph{IATSS research}, vol.~43, no.~4, pp. 244--252, 2019.

\bibitem{autocar_re1}
Z.~Guo, Y.~Huang, X.~Hu, H.~Wei, and B.~Zhao, ``A survey on deep learning based approaches for scene understanding in autonomous driving,'' \emph{Electronics}, vol.~10, no.~4, p. 471, 2021.

\bibitem{AutoMisty}
X.~Wang, L.~Dong, S.~Rangasrinivasan, I.~Nwogu, S.~Setlur, and V.~Govindaraju, ``Automisty: A multi-agent llm framework for automated code generation in the misty social robot,'' \emph{arXiv preprint arXiv:2503.06791}, 2025.

\bibitem{AI-MT}
E.~Baek, D.~Kwon, and J.~Kim, ``A multi-neural network acceleration architecture,'' in \emph{2020 ACM/IEEE 47th Annual International Symposium on Computer Architecture (ISCA)}.\hskip 1em plus 0.5em minus 0.4em\relax IEEE, 2020, pp. 940--953.

\bibitem{NB-SMT}
G.~Shomron and U.~Weiser, ``Non-blocking simultaneous multithreading: Embracing the resiliency of deep neural networks,'' in \emph{2020 53rd Annual IEEE/ACM International Symposium on Microarchitecture (MICRO)}.\hskip 1em plus 0.5em minus 0.4em\relax IEEE, 2020, pp. 256--269.

\bibitem{HDA}
H.~Kwon, L.~Lai, M.~Pellauer, T.~Krishna, Y.-H. Chen, and V.~Chandra, ``Heterogeneous dataflow accelerators for multi-dnn workloads,'' in \emph{2021 IEEE International Symposium on High-Performance Computer Architecture (HPCA)}.\hskip 1em plus 0.5em minus 0.4em\relax IEEE, 2021, pp. 71--83.

\bibitem{magma}
S.-C. Kao and T.~Krishna, ``Magma: An optimization framework for mapping multiple dnns on multiple accelerator cores,'' in \emph{2022 IEEE International Symposium on High-Performance Computer Architecture (HPCA)}.\hskip 1em plus 0.5em minus 0.4em\relax IEEE, 2022, pp. 814--830.

\bibitem{PREMA}
Y.~Choi and M.~Rhu, ``Prema: A predictive multi-task scheduling algorithm for preemptible neural processing units,'' in \emph{2020 IEEE International Symposium on High Performance Computer Architecture (HPCA)}.\hskip 1em plus 0.5em minus 0.4em\relax IEEE, 2020, pp. 220--233.

\bibitem{planaria}
S.~Ghodrati, B.~H. Ahn, J.~K. Kim, S.~Kinzer, B.~R. Yatham, N.~Alla, H.~Sharma, M.~Alian, E.~Ebrahimi, N.~S. Kim \emph{et~al.}, ``Planaria: Dynamic architecture fission for spatial multi-tenant acceleration of deep neural networks,'' in \emph{2020 53rd Annual IEEE/ACM International Symposium on Microarchitecture (MICRO)}.\hskip 1em plus 0.5em minus 0.4em\relax IEEE, 2020, pp. 681--697.

\bibitem{CD-MSA}
C.~Wang, Y.~Bai, and D.~Sun, ``Cd-msa: cooperative and deadline-aware scheduling for efficient multi-tenancy on dnn accelerators,'' \emph{IEEE Transactions on Parallel and Distributed Systems}, vol.~34, no.~7, pp. 2091--2106, 2023.

\bibitem{HASP}
H.~Li, S.~Ma, T.~Wang, W.~Zhang, G.~Wang, C.~Song, H.~Qu, J.~Lin, C.~Ma, J.~Pei \emph{et~al.}, ``Hasp: Hierarchical asynchronous parallelism for multi-nn tasks,'' \emph{IEEE Transactions on Computers}, vol.~73, no.~2, pp. 366--379, 2023.

\bibitem{MARS}
G.~Shen, J.~Zhao, Z.~Wang, Z.~Lin, W.~Ding, C.~Wu, Q.~Chen, and M.~Guo, ``Mars: Exploiting multi-level parallelism for dnn workloads on adaptive multi-accelerator systems,'' in \emph{2023 60th ACM/IEEE Design Automation Conference (DAC)}.\hskip 1em plus 0.5em minus 0.4em\relax IEEE, 2023, pp. 1--6.

\bibitem{moca_cacti}
S.~Kim, H.~Genc, V.~V. Nikiforov, K.~Asanovi{\'c}, B.~Nikoli{\'c}, and Y.~S. Shao, ``Moca: Memory-centric, adaptive execution for multi-tenant deep neural networks,'' in \emph{2023 IEEE International Symposium on High-Performance Computer Architecture (HPCA)}.\hskip 1em plus 0.5em minus 0.4em\relax IEEE, 2023, pp. 828--841.

\bibitem{zhao2025sparsemapsparsetensoraccelerator}
\BIBentryALTinterwordspacing
B.~Zhao, H.~Zhai, Z.~Yuan, H.~Liu, T.~Xia, W.~Zhao, and P.~Ren, ``Sparsemap: A sparse tensor accelerator framework based on evolution strategy,'' 2025. [Online]. Available: \url{https://arxiv.org/abs/2508.12906}
\BIBentrySTDinterwordspacing

\bibitem{UPot}
T.~Xia, B.~Zhao, J.~Ma, G.~Fu, W.~Zhao, N.~Zheng, and P.~Ren, ``An energy-and-area-efficient cnn accelerator for universal powers-of-two quantization,'' \emph{IEEE Transactions on Circuits and Systems I: Regular Papers}, 2022.

\bibitem{arvr_20ms}
Y.~Wang and J.~Zhao, ``A survey of mobile edge computing for the metaverse: Architectures, applications, and challenges,'' \emph{arXiv preprint arXiv:2212.00481}, 2022.

\bibitem{Llama-3}
A.~Dubey, A.~Jauhri, A.~Pandey, A.~Kadian, A.~Al-Dahle, A.~Letman, A.~Mathur, A.~Schelten, A.~Yang, A.~Fan \emph{et~al.}, ``The llama 3 herd of models,'' \emph{CoRR}, 2024.

\bibitem{deepseek}
X.~Bi, D.~Chen, G.~Chen, S.~Chen, D.~Dai, C.~Deng, H.~Ding, K.~Dong, Q.~Du, Z.~Fu \emph{et~al.}, ``Deepseek llm: Scaling open-source language models with longtermism,'' \emph{arXiv preprint arXiv:2401.02954}, 2024.

\bibitem{qwen}
J.~Bai, S.~Bai, Y.~Chu, Z.~Cui, K.~Dang, X.~Deng, Y.~Fan, W.~Ge, Y.~Han, F.~Huang \emph{et~al.}, ``Qwen technical report,'' \emph{arXiv preprint arXiv:2309.16609}, 2023.

\bibitem{graph_isomorphism}
G.~Bouritsas, F.~Frasca, S.~Zafeiriou, and M.~M. Bronstein, ``Improving graph neural network expressivity via subgraph isomorphism counting,'' \emph{IEEE Transactions on Pattern Analysis and Machine Intelligence}, vol.~45, no.~1, pp. 657--668, 2023.

\bibitem{lu2021sanger}
L.~Lu, Y.~Jin, H.~Bi, Z.~Luo, P.~Li, T.~Wang, and Y.~Liang, ``Sanger: A co-design framework for enabling sparse attention using reconfigurable architecture,'' in \emph{MICRO-54: 54th Annual IEEE/ACM International Symposium on Microarchitecture}, 2021, pp. 977--991.

\bibitem{period_survey}
R.~I. Davis and A.~Burns, ``A survey of hard real-time scheduling for multiprocessor systems,'' \emph{ACM computing surveys (CSUR)}, vol.~43, no.~4, pp. 1--44, 2011.

\bibitem{Herald}
H.~Kwon, L.~Lai, M.~Pellauer, T.~Krishna, Y.-H. Chen, and V.~Chandra, ``Heterogeneous dataflow accelerators for multi-dnn workloads,'' in \emph{2021 IEEE International Symposium on High-Performance Computer Architecture (HPCA)}.\hskip 1em plus 0.5em minus 0.4em\relax IEEE, 2021, pp. 71--83.

\bibitem{CV1}
G.~F. Reed, F.~Lynn, and B.~D. Meade, ``Use of coefficient of variation in assessing variability of quantitative assays,'' \emph{Clinical and Vaccine Immunology}, vol.~9, no.~6, pp. 1235--1239, 2002.

\bibitem{CV2}
H.~Abdi, ``Coefficient of variation,'' \emph{Encyclopedia of research design}, vol.~1, no.~5, pp. 169--171, 2010.

\bibitem{FreePDK45}
J.~E. Stine, I.~Castellanos, M.~Wood, J.~Henson, F.~Love, W.~R. Davis, P.~D. Franzon, M.~Bucher, S.~Basavarajaiah, J.~Oh, and R.~Jenkal, ``Freepdk: An open-source variation-aware design kit,'' in \emph{2007 IEEE International Conference on Microelectronic Systems Education (MSE'07)}, 2007, pp. 173--174.

\bibitem{McPAT}
S.~Li, J.~H. Ahn, R.~D. Strong, J.~B. Brockman, D.~M. Tullsen, and N.~P. Jouppi, ``Mcpat: An integrated power, area, and timing modeling framework for multicore and manycore architectures,'' in \emph{Proceedings of the 42nd annual ieee/acm international symposium on microarchitecture}, 2009, pp. 469--480.

\bibitem{Mobilenetv2}
M.~Sandler, A.~Howard, M.~Zhu, A.~Zhmoginov, and L.-C. Chen, ``Mobilenetv2: Inverted residuals and linear bottlenecks,'' in \emph{Proceedings of the IEEE conference on computer vision and pattern recognition}, 2018, pp. 4510--4520.

\bibitem{resnet50}
K.~He, X.~Zhang, S.~Ren, and J.~Sun, ``Identity mappings in deep residual networks,'' in \emph{Computer Vision--ECCV 2016: 14th European Conference, Amsterdam, The Netherlands, October 11--14, 2016, Proceedings, Part IV 14}.\hskip 1em plus 0.5em minus 0.4em\relax Springer, 2016, pp. 630--645.

\bibitem{Efficientnet}
M.~Tan and Q.~Le, ``Efficientnet: Rethinking model scaling for convolutional neural networks,'' in \emph{International conference on machine learning}.\hskip 1em plus 0.5em minus 0.4em\relax PMLR, 2019, pp. 6105--6114.

\bibitem{atomic_DAG}
S.~Zheng, X.~Zhang, L.~Liu, S.~Wei, and S.~Yin, ``Atomic dataflow based graph-level workload orchestration for scalable dnn accelerators,'' in \emph{2022 IEEE International Symposium on High-Performance Computer Architecture (HPCA)}.\hskip 1em plus 0.5em minus 0.4em\relax IEEE, 2022, pp. 475--489.

\bibitem{U-net}
O.~Ronneberger, P.~Fischer, and T.~Brox, ``U-net: Convolutional networks for biomedical image segmentation,'' in \emph{International Conference on Medical image computing and computer-assisted intervention}.\hskip 1em plus 0.5em minus 0.4em\relax Springer, 2015, pp. 234--241.

\bibitem{NASNet}
B.~Zoph, V.~Vasudevan, J.~Shlens, and Q.~V. Le, ``Learning transferable architectures for scalable image recognition,'' in \emph{Proceedings of the IEEE conference on computer vision and pattern recognition}, 2018, pp. 8697--8710.

\bibitem{PNASNet}
C.~Liu, B.~Zoph, M.~Neumann, J.~Shlens, W.~Hua, L.-J. Li, L.~Fei-Fei, A.~Yuille, J.~Huang, and K.~Murphy, ``Progressive neural architecture search,'' in \emph{Proceedings of the European conference on computer vision (ECCV)}, 2018, pp. 19--34.

\bibitem{MLPerf}
V.~J. Reddi, C.~Cheng, D.~Kanter, P.~Mattson, G.~Schmuelling, C.-J. Wu, B.~Anderson, M.~Breughe, M.~Charlebois, W.~Chou \emph{et~al.}, ``Mlperf inference benchmark,'' in \emph{2020 ACM/IEEE 47th Annual International Symposium on Computer Architecture (ISCA)}.\hskip 1em plus 0.5em minus 0.4em\relax IEEE, 2020, pp. 446--459.

\bibitem{xia2022energy}
T.~Xia, B.~Zhao, J.~Ma, G.~Fu, W.~Zhao, N.~Zheng, and P.~Ren, ``An energy-and-area-efficient cnn accelerator for universal powers-of-two quantization,'' \emph{IEEE Transactions on Circuits and Systems I: Regular Papers}, vol.~70, no.~3, pp. 1242--1255, 2022.

\end{thebibliography}
%

\bibliographystyle{IEEEtran}

\vfill

\end{document}